\documentclass[
aps,
pra,
superscriptaddress,
showpacs,
showkeys,
notitlepage,
twocolumn, 
numbers,
longbibliography,
nofootinbib,
floatfix]
{revtex4-2}

\usepackage[T1]{fontenc} 
\usepackage[utf8]{inputenc} 
\usepackage[english]{babel} 
\usepackage{graphicx, xcolor}
\usepackage{amsfonts, amssymb, amsmath, mathrsfs, calc, array, mathtools}
\usepackage[separate-uncertainty=true]{siunitx}
\usepackage{abraces}
\usepackage{bbm}
\PassOptionsToPackage{hyphens}{url}
\usepackage{hyperref}
\usepackage{url}

\usepackage{tikz}
\usetikzlibrary{arrows.meta}
\usepackage{enumitem}

\def\ontop#1#2{\setbox0\hbox{#2}\copy0\llap{\raise\ht0\hbox{#1}}}

\newcommand{\defeq}{\vcentcolon=}

\definecolor{darkblue}{rgb}{0,0,0.93} 
\definecolor{darkred}{rgb}{0.8,0,0} 
\definecolor{darkgreen}{rgb}{0,0.7,0} 

\hypersetup{
colorlinks=true, 
linkcolor=darkblue, 
citecolor=darkred, 
filecolor=darkblue, 
urlcolor=darkblue,
breaklinks=true
}

\makeatletter
\def\l@subsubsection#1#2{}
\makeatother

\begin{document}

\title{A single-particle framework for unitary lattice gauge theory in discrete time}

\author{Pablo Arnault}
\email{pablo.arnault@inria.fr}
\affiliation{Univ.\ Paris-Saclay,   CNRS, ENS Paris-Saclay,  INRIA, Laboratoire Méthodes Formelles,  91190 Gif-sur-Yvette,  France}

\author{Christopher Cedzich}
\email{cedzich@hhu.de}
\affiliation{Quantum Technology Group, Heinrich Heine Universit\"at D\"usseldorf, Universit\"atsstr. 1, 40225 D\"usseldorf, Germany}

\begin{abstract}
We construct a real-time lattice-gauge-theory-type action for a spin-1/2 matter field of a single particle on a $(1+1)$-dimensional spacetime lattice.
The framework is based on a discrete-time quantum walk, and  is hence inherently unitary and strictly local, i.e., transition amplitudes exactly vanish outside of a lightcone on the lattice.
We then provide a lattice Noether's theorem for internal symmetries of this action. We further couple this action to an electromagnetic
field by a minimal substitution on the lattice.
Finally, we suggest a real-time lattice-gauge-theory-type action for the electromagnetic field in arbitrary spacetime dimensions, and derive its classical equations of motion, which are lattice versions of Maxwell's equations.
\end{abstract}

\maketitle

\section{Introduction}

Lattice gauge theory (LGT) is a framework used to tackle the non-perturbative regimes of quantum field theories (QFTs) \cite{book_Rothe}.
The basic idea is to formulate the gauge theory under study on a lattice, which furnishes an in-built ultraviolet regulator.
LGT was initially introduced by Wilson \cite{Wilson74} as a framework accounting successfully for quark confinement; the gauge theory in question was (a simplified version of) quantum chromodynamics (QCD).

There are two fundamental types of LGTs.
The traditional one is ``Lagrangian LGT'' \cite{Wilson74}, in which the main object of the theory is an action in discrete spacetime.
Lagrangian LGT treats time and space on the same footing, and is in this sense close to Einstein's theory of relativity.
However, in Lagangian LGT unitarity is not manifest -- it has to be proven and does not always hold, see, e.g., Ref.\ \cite{Hernandez2011}.
The second type of LGT is ``Hamiltonian LGT'' \cite{WK74,KogSuss75a}, in which the main object of the theory is a Hamiltonian on a spatial lattice while time is usually kept continuous.
The pros and cons are therefore reversed with respect to Lagrangian LGT: Hamiltonian LGTs are unitary by construction, but space and time are often treated differently. 
These differences between Lagrangian and Hamiltonian LGT are actually already present in the continuum: Lagrangian LGT comes from the path-integral approach to QFT, in which Lorentz covariance is manifest but unitarity is not, whereas Hamiltonian LGT comes from the canonical approach to QFT, which is manifestly unitary but not manifestly Lorentz covariant \cite{Weinberg_QFT1}.

The standard tool of lattice QCD are Monte-Carlo (MC) simulations, which are usually carried out in the Lagrangian formulation with a Wick-rotated spacetime, i.e., in Euclidean spacetime with imaginary time \cite{book_Rothe}.
MC simulations have been successfully used to determine (i) equilibrium properties of QCD, such as the masses of quarks and stable hadrons, hadronic structure-related quantities and non-zero-temperature properties \cite{BCCJplus17}, and (ii) certain non-equilibrium, i.e., beyond-ground-states properties \cite{Lang2008, Aoki2021}.
Despite its successes, standard LGT with MC simulations fails to give results in several cases, in particular in parameter regimes in which MC simulations encounter a so-called sign problem \cite{BCCJplus17}, which can almost certainly not be solved efficiently by (even the most powerful \cite{BCDEplus2018}) classical computers with traditional techniques \cite{TW2005}.
Several techniques exist to overcome the sign problem for certain simplified models (see references in Ref. \cite{BCCJplus17}), including tensor-network approaches \cite{BCCJplus17}, which use a real-time Hamiltonian formulation, where we are going to define just below what ``real-time'' means.
 
Let us call ``real-time LGT'' any LGT framework in which spacetime is not Wick-rotated.
Because time is kept real, all such frameworks are expected to be particularly suited to explore the dynamics of the gauge theory in question \cite{Preskill2018}. Thereby, real-time LGT is one privileged approach to computing non-equilibrium properties of gauge theories. 
Lately, real-time LGT has entered a new era with the advent of quantum simulation \cite{Feynman1982} and quantum computation \cite{book_nielsen_chuang_2010}, which reduce exponentially the cost in simulating many-body quantum systems. 
As above, there are two types of real-time LGTs: (i) real-time Hamiltonian LGT, where time is either kept continuous (with the perspective of analog quantum simulations \cite{ZR2011, BDMRplus2012, Wiese2013, ZCR2015}) or discretized (with the perspective of digital quantum simulations \cite{WMLZplus2010, TCZL2013} and quantum algorithms \cite{Byrnes2006, JLP2012, JLP2014, JLP2014b, BRSS2015, MPSW2015, JKLP2018, ABLL2019, MGJ2019}), and (ii) real-time Lagrangian LGT, where time is discretized but spacetime is not Wick-rotated \cite{KW21}.
Our paper combines both perspectives.
Reviews with different focuses on the general topic of quantum-information inspired methods for QFTs can be found in Refs.\ \cite{DM2016, Preskill2018, Banuls2020, BC2020, Aidelsburger2021, Zohar2021, KRS2022}.
Several proof-of-principle experiments of quantum simulations of the dynamics of LGTs have already been realized \cite{Martinez16} (see also references in Ref.\ \cite{Zohar2021}).

Among the real-time Hamiltonian-LGT approaches in discrete time, there is one, often not known under the name of LGT, which consists in using quantum cellular automata (QCA) \cite{Arrighi2019, Farrelly2020,schumacher2004reversible}.
QCA are by-construction unitary evolution operators in discrete spacetime that are \emph{strictly local}, i.e., there is an in-built strict ``relativistic'' lightcone at the discrete-spacetime level. 
Results in this field of ``QCA LGT'' are still preliminary, but promising \cite{DDV87, BDAPplus2018, ABF20, SADM2022, EDMMplus22, Farrelly15, FS2020, Yepez2016}.
Let us stress this important point: in the usual discrete-time Hamiltonian-LGT approaches, locality enters merely in the form of an effective lightcone \cite{CJWW} related to Lieb-Robinson bounds, whereas QCA are strictly local by construction.
At the level of classical fields, i.e., in the one-particle sector, QCA reduce to so-called discrete-time quantum walks (DQWs) \cite{Vogts09}. 
Understanding DQWs as the building blocks of QCA, one can therefore expect that the field of QCA LGT will benefit in the near future from the numerous results that exist (i) in the one-particle sector, in the free case \cite{BB94a, Farrelly2014a, Bisio2015}, with couplings to Abelian \cite{AD16b, MMAMP18, CGWW18} and non-Abelian \cite{AMBD16} gauge fields, and with curved spacetimes \cite{DMD13b, DMD14, AD17, Arrighi_curved_1D_15, AF17, Arnault2017, ADMMMplus2019}, but also (ii) in the multiparticle free case \cite{Farrelly2014b, DAP16}.

In this manuscript, we further complete the list of achievements in the one-particle sector with the following results:
\begin{enumerate}[label=(\arabic*)]
    \item We construct a real-time LGT-type action for spin-1/2 matter fields on a $(1+1)$-dimensional spacetime lattice, that is based on a DQW. This ``DQW action'' is therefore by-construction unitary -- i.e., it delivers unitary equations of motion (EOMs) --, and it treats time and space on the same footing.
    \item We provide a lattice Noether's theorem for the internal symmetries of the DQW action. Applying this theorem to the global U(1) symmetry of the DQW action, we find a U(1)-charge current conserved on the lattice.
    \item We place the particle into an  Abelian U(1) gauge field by applying a lattice minimal-coupling scheme to the DQW action.
    \item For this Abelian U(1) gauge field which generates an electromagnetic field, we suggest a real-time LGT-type action in arbitrary spacetime dimensions, from which we derive the classical EOMs of the gauge field, which are lattice versions of Maxwell’s equations.
\end{enumerate}

Let us give more details on these achievements.
We bring together in a unified framework elements from the field of QCA with elements from Lagrangian LGT, and this for spin-1/2 matter fields. 
In this sense, we extend Ref.\ \cite{FS2020} which considers scalar fields. 
All is done at the level of classical fields, i.e., the fields are not quantized\footnote{In Ref.\ \cite{FS2020} the scalar fields are quantum.}.
Hence, throughout this paper we use ``fermionic'' synonymously to ``spin 1/2''.

The definition of a real-time LGT-type action $S_{\text{DQW}}$ based on a DQW is the first stepping stone in this paper. 
To the best of our knowledge, the only paper that suggests a discrete-spacetime action for a spin-1/2 matter field based on DQWs is Ref.\ \cite{Debbasch2019a}.
However, the action in Ref.\ \cite{Debbasch2019a} does not relate nicely to usual LGT actions in the sense that it is based on a \emph{one-step} equation of motion (EOM) for the matter field $\psi$, i.e., an EOM of the type $\psi_{j+1} = \mathcal{W}\psi_j$,  where $j\in \mathbb{Z}$ labels discrete time and $\mathcal{W}$ is a unitary evolution operator. 
In contrast, in LGT one usually constructs EOMs with symmetric finite differences; these \emph{two-step} EOMs involve a field $\psi$ at three subsequent time instants  $\psi_{j-1}$, $\psi_j$ and $\psi_{j+1}$, and therefore require two initial conditions rather than just a single one.
Whether one- or two-step, EOMs in LGT usually do not preserve the unitarity of the continuum model.
Building on the construction of Ref.\ \cite{Arnault2022}, we remedy this lack by providing a \emph{unitary} real-time action, that is extremely similar to standard LGT actions in the sense that it is based on a two-step EOM, while being associated with a DQW (hence its unitarity) with a one-step EOM.

The remainder of this paper is organized as follows.
We begin with discussing one- and two-step EOMs for spin-1/2 particles in discrete spacetime in Sec.\ \ref{sec:eom}, which leads us to the definition of the corresponding DQW action $S_{\text{DQW}}$ in Sec.\ \ref{subsec:action_matter_field}.
In Sec.\ \ref{sec:Noether} we prove a lattice Noether's theorem for internal symmetries of a generic real-time action for spin-1/2 particles.
In Sec.\ \ref{sec:gauging} we couple $S_{\text{DQW}}$ to an Abelian U(1)
gauge field  via a lattice version of minimal substitution.
In the last section, Sec.\ \ref{sec:action_gauge_field}, we suggest a real-time LGT-type action for this Abelian U(1) gauge field, in arbitrary spacetime dimensions, and derive the corresponding classical EOMs, which are lattice versions of Maxwell's equations.

\section{DQW-based LGT-type action for a classical matter field}
\label{sec:eom}

\subsection{The continuum equation of motion}

Consider in $1+1$ dimensions a relativistic classical matter field $\psi$, with internal components $\psi^a$ where $a=1,...,N$ for some $N  \in \mathbb{N}$.
The dynamics of $\psi$ is described by the Dirac equation
\begin{equation}
\label{eq:Dirac_eq}
(\mathrm{i}\gamma^{\mu}\partial_{\mu} - m)\psi = 0 \, ,
\end{equation}
where the $\gamma_{\mu}$, $\mu=0,1$, act on the internal Hilbert space, and satisfy the Clifford-algebra relations
\begin{equation}
\label{eq:Clifford_algebra}
\{\gamma^{\mu},\gamma^{\nu} \} = 2\eta^{\mu\nu} \, ,
\end{equation}
with $[\eta^{\mu\nu}] \defeq \text{diag}(1,-1)$, and where on the right-hand side of Eq.\ \eqref{eq:Clifford_algebra} we omitted for brevity the identity on the internal Hilbert space. 

The Dirac equation \eqref{eq:Dirac_eq} can be rewritten in the form of a Schrödinger equation,
\begin{equation}
\label{eq:Dirac_eq_Hamiltonian}
\mathrm{i}\partial_t \psi = \mathcal{H} \psi \, .
\end{equation}
Here, $\partial_t \equiv \partial_0$ is the partial derivative with respect to time, and we have introduced the Dirac Hamiltonian
\begin{equation}
 \mathcal{H}  \defeq \alpha^1 (-\mathrm{i}\partial_{1}) + m \alpha^0 \, ,
\end{equation}
where $\partial_1 \equiv \partial_{x^1}$ is the partial derivative with respect to the spatial position $x^1$, and where we have introduced the operators 
\begin{subequations}
\begin{align}
\alpha^0 &\defeq \gamma^0 \\
\alpha^1 &\defeq \gamma^0 \gamma^1 \, ,
\end{align}
\end{subequations}
which satisfy the relations
\begin{equation}
\label{eq:alpha_algebra}
\{\alpha^{\mu},\alpha^{\nu} \} = 2\delta^{\mu\nu} \, ,
\end{equation}
where $[\delta^{\mu\nu}] \defeq \text{diag}(1,1)$.

In one spatial dimension, it is enough to consider an internal Hilbert space of dimension 2 to find a pair of alpha matrices $([{(\alpha^0)}^{a}_{b}], [{(\alpha^1)}^{a}_{b}])$ that satisfy Eq.\ \eqref{eq:alpha_algebra}.
In that case, the index $a$ of the internal Hilbert space of $\psi$ belongs to $\{1,2\}$.
Unless otherwise mentioned we will work with the abstract objects $\alpha^0$ and $\alpha^1$ rather than their matrix representations, such that in particular the dimension of the latter will not play a role.
Hence, the notation ``$\psi$'' is abstract in the internal Hilbert space but ``concrete'' in the position Hilbert space.

\subsection{The naive discretization}

We introduce the $(1+1)$-dimensional spacetime lattice $\mathbb Z\times\mathbb Z$ and label its sites by the multi-index $n\equiv(j,p)$. Denoting the spacetime-lattice spacing by $\epsilon$, we take $j$ to label time, i.e., we set $t\equiv j\epsilon$, and $p$ to label position, i.e., $x^1\equiv p\epsilon$. We also define $x\equiv (t,x^1)$, which we use flexibly in the continuum as well as in the discrete, in which case $x \equiv \epsilon n $. Moreover, we write
\begin{equation}
\psi_n \defeq \psi(\epsilon n) \,
\end{equation}
for the field $\psi$ evaluated at the lattice site $n$.

\subsubsection{Lattice derivatives}

A standard way \cite{book_Rothe} of discretizing Eq.\ \eqref{eq:Dirac_eq_Hamiltonian} in space and time while preserving the Hermiticity of the operators $\mathrm{i}\partial_{\mu}$, $\mu=0,1$, and thus that of the Hamiltonian, is to use \emph{symmetric finite differences}, i.e., to perform the substitution
\begin{equation}\label{eq:discretize}
\mathrm{i}\partial_{\mu} \longrightarrow \mathrm{i}d_\mu:=\frac{\mathrm{i}}{2\epsilon}\left(\mathcal{T}_{\mu}^{-1} - \mathcal{T}_{\mu} \right).
\end{equation} 
As above, $\epsilon$ is the spacetime-lattice spacing, and $\mathcal{T}_{\mu}$ is the translation operator in direction $\mu$, i.e.,
\begin{equation}
\label{eq:translation_operators}
(\mathcal{T}_{\mu} \psi)_n = \psi_{n - \hat{\mu}} \, ,
\end{equation}
where $\hat{\mu}$ is the unit vector in direction $\mu$, compare with Ref.\ \cite[Eq.\ (33)]{CGWW18}.
Clearly, $\frac{\mathrm{i}}{2\epsilon}\left(\mathcal{T}_{\mu}^{-1} - \mathcal{T}_{\mu} \right)$ is Hermitian because $\mathcal{T}_{\mu}$ is unitary. Moreover, in the continuum limit $\epsilon\to0$ we have $ d_\mu\to\partial_\mu$.

For later use we also introduce the left and right lattice derivatives
\begin{subequations}
\label{eq:LR-derivs}
\begin{align}
\label{eq:L-deriv}
    d_{\mu}^L &\defeq  \frac{1}{\epsilon} \left( 1 - \mathcal{T}_{\mu} \right)\\
    d_{\mu}^R &\defeq \frac{1}{\epsilon} \left(  \mathcal{T}_{\mu}^{-1}-1 \right)\, ,
    \label{eq:R-deriv}
\end{align}
\end{subequations}
such that $d_\mu=(d_{\mu}^L+d_{\mu}^R)/2$.

\subsubsection{Standard-LGT scheme: naive fermions}

Discretizing the Dirac equation, Eq.\ \eqref{eq:Dirac_eq_Hamiltonian}, with the symmetric finite differences yields
\begin{equation}
\label{eq:discretization}
 \, \frac{\mathrm{i}}{2\epsilon}(\mathcal{T}_0^{-1}-\mathcal{T}_0)  \,  \psi = -\alpha^1   \,  \frac{\mathrm{i}}{2\epsilon}(\mathcal{T}_1^{-1}-\mathcal{T}_1) \psi + \alpha^0 m \psi \,.
\end{equation}
This equation can be rewritten as
\begin{equation}
\label{eq:naive_fermions}
\mathrm{i} d_0 {\psi} = {\mathcal{H}}^{\text{LGT}} {\psi} \, ,
\end{equation}
a scheme which we call that of \emph{naive fermions}, where the lattice Hamiltonian is
\begin{equation}
\label{eq:naive_ham}
{\mathcal{H}}^{\text{LGT}} \defeq \alpha^1 (-\mathrm{i} d_1) +  {m} \alpha^0 \, .
\end{equation}
The associated action can be found in Ref.\ \cite[Sec.\ 4.1]{book_Rothe}.

A comment must be made at this stage: the lattice Hamiltonian of Eq.\ \eqref{eq:naive_ham} is usually introduced in Hamiltonian LGTs  in which time is kept continuous \cite{KogSuss75a, Susskind77a}, and in those frameworks the role of the Hamiltonian is the usual one, that is, it generates the time evolution. Here, we introduce ${\mathcal{H}}^{\text{LGT}}$ even though we are in discrete time, and we call it a ``Hamiltonian'' even though it does not generate the time evolution in the usual sense: time evolution is described instead by the lattice EOM of Eq.\ \eqref{eq:naive_fermions}.

\subsection{The DQW discretization}

The use of the symmetric lattice derivatives above ensures the Hermiticity of the lattice Hamiltonian ${\mathcal{H}}^{\text{LGT}}$.
Thus, if time was not discretized and time evolution was generated by this lattice Hamiltonian, the scheme would be unitary.
However, when also discretizing time, what about unitarity?

In textbook Lagrangian LGT, discrete-time formulations are usually carried out in Euclidean spacetime.
In such a framework, unitarity of the model is proven essentially by proving the positivity of the transfer operator\footnote{A transfer operator is any operator $\hat{T}$ which in the continuum limit coincides with $\exp(-\tau \hat{H})$, where $\hat{H}$ is the Hamiltonian of the system, but which at the discrete level could differ from merely exponentiating $\hat{H}$. In the discrete, the exact form of $\hat{T}$ is adjusted in order to facilitate the computation of the transition amplitudes $\langle n | \hat{T} | n' \rangle$. For example, for a point particle of mass $m$ in a potential $V$, a possible suitable definition for the transfer operator is $T \defeq \exp(-\tau V(\hat{x})/2) \exp(-\tau \hat{p}^2/(2m)) \exp(-\tau V(\hat{x})/2) $, where $\hat{x}$ is the position operator, and $\hat{p}$ is the momentum operator. The naive transfer operator is simply the Euclidean version, $\exp(-\tau \hat{H})$, of the one-step evolution operator $\exp(-\mathrm{i} t \hat{H})$.} of the system, or alternatively by proving the so-called Osterwalder–Schrader reflection-positivity condition \cite{Hernandez2011}.

These properties are far from straightforward to establish: for example, the positivity of the transfer matrix for lattice fermions has only been proven for Wilson fermions with Wilson parameter $r=1$, and there is no proof for naive fermions, which have $r=0$ \cite{Hernandez2011}, i.e., there is no proof of the positivity of the transfer operator for the scheme of Eq.\ \eqref{eq:naive_fermions}.
Based on \ref{subsubsec:modif_naive} and \ref{subsubsec:underlying} below, we believe that this scheme is simply not unitary\footnote{That is to say, more precisely, there exists no underlying unitary one-step scheme that generates naive fermions, see below.}.

\subsubsection{Modification of naive fermions for unitarity in discrete time}
\label{subsubsec:modif_naive}

First, notice that the scheme of naive fermions, Eq.\ \eqref{eq:naive_fermions}, is -- because of the use of a symmetric lattice derivative for time -- a \emph{two-step} scheme, i.e., it needs two initial conditions ${\psi}_{j=0}$ and $\psi_{j=1}$. 
As in Ref.\ \cite{Arnault2022}, let us perform in Eq.\ \eqref{eq:naive_fermions} the following substitution,
\begin{equation}\label{eq:tildealpha}
\alpha^{\mu} \longrightarrow \tilde{\alpha}^{\mu}\defeq\mu_{\epsilon} \alpha^{\mu} \, , \ \ \mu = 0,1 \, ,
\end{equation}
where the prefactor
\begin{equation}
\label{eq:nu}
\mu_{\epsilon} \defeq \frac{1}{\sqrt{1-  (\epsilon {m})^2}} \, ,
\end{equation}
is positive for $\epsilon m$ small enough.
This substitution \eqref{eq:tildealpha} leads to a new two-step scheme which we call \emph{unitary fermions},
\begin{equation}
\label{eq:DQW_fermions}
\mathrm{i}  d_0 {\psi} = {\mathcal{H}}^{\text{DQW}} {\psi} \, ,
\end{equation}
where the lattice Hamiltonian is now
\begin{equation}
\label{eq:HDQW}
{\mathcal{H}}^{\text{DQW}} \defeq \tilde{\alpha}^1 (-\mathrm{i}  d_1) +  m \tilde{\alpha}^0 \, .
\end{equation}
We are going to see in the next subsection that this new two-step scheme, Eq.\ \eqref{eq:DQW_fermions},
is equivalent to a \emph{unitary} one-step scheme,
\begin{equation}
\label{eq:one-step scheme}
\mathcal T_0^{-1}\psi_j\equiv{\psi}_{j+1} ={\mathcal{W}} {\psi}_j \, ,
\end{equation}
with properly chosen one-step unitary evolution operator $\mathcal W$ and provided that the second initial condition ${\psi}_{j=1}$ is given precisely by this unitary one-step scheme, that is,
\begin{equation}
{\psi}_{j=1} =  {\mathcal{W}} {\psi}_{j=0} \, .
\end{equation}
The two schemes are equivalent in the usual sense: if $\psi_j$ satisfies the one-step scheme, then it satisfies the two-step scheme, and vice versa. 
Note that this equivalence has already been proven in Ref.\ \cite{Arnault2022}. For the convenience of the reader we review it in the next subsection.

\subsubsection{The ``underlying'' unitary one-step scheme}
\label{subsubsec:underlying}

Let us explicate how the two-step scheme of Eq.\ \eqref{eq:DQW_fermions} can be obtained from 
a one-step scheme of the type of Eq.\ \eqref{eq:one-step scheme}.
Since the field $\psi$ in Eq.\ \eqref{eq:DQW_fermions} satisfies a first-order difference equation in position, we consider the unitary operator
\begin{equation}
\label{eq:walk_operator}
{\mathcal{W}} \defeq W_{-1} \mathcal{T}^{-1}_1 + W_{+1}  \mathcal{T}_1 + W_0\mathbbm1 \, ,
\end{equation}
where the \emph{jump operators} $W_i$, $i=-1,0,1$, act on the internal Hilbert space of ${\psi}$, and $\mathcal{T}_1$ is the translation operator defined in Eq.\ \eqref{eq:translation_operators}. 
We henceforth call $\mathcal W$ a \emph{walk operator} in accordance with the literature: it is a unitary evolution operator by one time step which is strictly local.

We define the following \emph{transport operators},
\begin{subequations}
\begin{align}
B_\pm &\defeq W_1 \pm W_{-1} \\
M &\defeq \sum_{i=-1,0,+1} W_i = B_+ + W_0 \, ,
\end{align}
\end{subequations}
which encode transport properties of the scheme and of its continuum limit, as we are going to see below.

Under which conditions on $B_+$, $B_-$ and $M$, can we obtain the two-step scheme of Eq.\ \eqref{eq:DQW_fermions} from the one-step scheme of Eq.\ \eqref{eq:one-step scheme}?
Notice first that if ${\psi}$ satisfies Eq.\ \eqref{eq:one-step scheme}, then it also satisfies the following two-step scheme,
\begin{equation}
\label{eq:generic_two-step_scheme}
\mathrm{i}d_0 {\psi} = {\mathcal{H}}_{Q} {\psi} \, ,
\end{equation}
where
\begin{subequations}
\begin{align}
\epsilon { \mathcal{H}}_{Q} & \defeq \frac{\mathrm{i}}{2} \left(\mathcal{W} - \mathcal{W}^{\dag}\right) \label{eq:local_Ham} \\
&= A^1 (-\mathrm{i} \epsilon d_1) + \frac{r}{2} Q (-\mathcal{L}) + \epsilon m {A}^0 \, ,
\end{align}
\end{subequations}
where we have (i) introduced the lattice Laplacian
\begin{equation}
\mathcal{L} \defeq \mathcal{T}^{-1}_1 + \mathcal{T}_1 - 2 \, ,
\end{equation} 
(ii) forced the appearance of the Wilson parameter $r$ and the ``mass'' $m$, and (iii) introduced the following operators acting on the internal Hilbert space,
\begin{subequations}
\begin{align}
{A}^0 &\defeq \frac{\mathrm{i}}{2 \epsilon m}(M-M^{\dag}) \\
A^1 &\defeq \frac{1}{2} \left( B_- + B_-^{\dag} \right) \\
Q &\defeq - \frac{\mathrm{i}}{2r}(B_+-B_+^{\dag})  \, .
\end{align}
\end{subequations}
Notice that the two-step local Hamiltonian of a DQW, defined in Eq.\ \eqref{eq:local_Ham}, was already introduced in Ref.\ \cite{APP20} before appearing in Ref.\ \cite{Arnault2022}.

Now, for $\epsilon \mathcal{H}_Q$ to equal $\epsilon \mathcal{H}^{\text{DQW}}$, we can choose
\begin{subequations}
\label{eqs:operators}
\begin{align}
{A}^0 &= \tilde{\alpha}^0 \\
A^1 &= \tilde{\alpha}^1 \\
Q &= 0 \, ,
\end{align}
\end{subequations}
with $\tilde{\alpha}^{\mu}$ from Eq.\ \eqref{eq:tildealpha}.
For the first two equations of Eqs.\ \eqref{eqs:operators} to hold, we can choose
\begin{subequations}
\label{eq:choices}
\begin{align}
M &= \mu_{\epsilon} (1	- \mathrm{i} \epsilon m \alpha^0) \\
B_- &= \tilde{\alpha}^1 \, , \label{eq:B}
\end{align}
\end{subequations}
and for $Q$ to vanish we must choose
\begin{equation}
\label{eq:V}
B_+=B_+^{\dag} \, .
\end{equation}
Note that this implies
\begin{equation}
    W_1-W_1^{\dag}=-(W_{-1}-W_{-1}^{\dag}) \, .
\end{equation}

It is easy to show that the choices of Eqs.\ \eqref{eq:choices} are compatible with the unitarity constraints involving solely $B_-$ and $M$ that result from $\mathcal{W}^{\dag}\mathcal{W} = \mathbf 1 = \mathcal{W} \mathcal{W}^{\dag} $, see \cite[Appendix A]{Arnault2022}.
A choice for $B_+$ that satisfies Eq.\ \eqref{eq:V} and that is compatible with all unitarity constraints involving $B_+$ \cite[Appendix A]{Arnault2022} is ``simply'' 
\begin{equation}
\label{eq:choiceV}
B_+ = \mu_{\epsilon} \, .
\end{equation}
In the end, we have found a unitary one-step scheme, namely, Eq.\ \eqref{eq:one-step scheme} with the choices of Eqs.\ \eqref{eq:choices} and \eqref{eq:choiceV}, which generates the two-step scheme of \emph{unitary  fermions}.

Let us as a sum-up explicitly write this scheme of \emph{one-step unitary fermions}:
\begin{equation}
\label{eq:sum-up}
    \psi_{j+1} = \mathcal{W}_{\text{Dirac}}\psi_j \, ,
\end{equation}
with
\begin{equation}
    \mathcal{W}_{\text{Dirac}} \defeq \mu_{\epsilon} \Big[ \frac{1}{2}(1-\alpha^1) \mathcal{T}^{-1} + \frac{1}{2}(1+\alpha^1) \mathcal{T} -\mathrm{i}\epsilon m \alpha^0  \Big] \, .
\end{equation}

\subsection{The lattice action of the matter field}
\label{subsec:action_matter_field}

The only difference between naive fermions and unitary  fermions is the prefactor $\mu_{\epsilon}$ appearing in the modified $\alpha$ operators and hence also in the following modified $\gamma$ operators,
\begin{subequations}
\label{eqs:tildegammas}
\begin{align}
\tilde{\gamma}^0 &\defeq (\tilde{\alpha}^0)^{-1} = \frac{\alpha^0}{\mu_{\epsilon}} = \frac{\gamma^0}{\mu_{\epsilon}}  \\
\tilde{\gamma}^1 &\defeq (\tilde{\alpha}^0)^{-1}  \tilde{\alpha}^1 = \gamma^1 \, .
\end{align}
\end{subequations}

It turns out that one can obtain a valid lattice action for unitary  fermions by performing the substitution $\gamma^{\mu} \longrightarrow \tilde{\gamma}^{\mu}$ in the, e.g., \emph{asymmetric} lattice action of naive fermions.
This results in 
\begin{equation}
\label{eq:action}
S_{\text{DQW}}^{\text{asym.}}  \defeq \epsilon^2 \sum_n \bar{\bar{\psi}}_n \left[ \left( \mathrm{i} \tilde{\gamma}^{\mu} d_{\mu} - {m} \right) \psi \right]_n \, ,
\end{equation}
with
\begin{equation}
\label{eq:psibar}
\bar{\bar{\psi}}_n \defeq \psi^{\dag}_n (\tilde{\gamma}^0)^{-1} \,  .
\end{equation}
In Appendix \ref{app:variational_principle}, we show how extremalizing this action, Eq.\ \eqref{eq:action}, indeed yields the correct EOM of unitary  fermions, Eq.\ \eqref{eq:DQW_fermions}.

We call the action in Eq.\ \eqref{eq:action} ``asymmetric'', since the lattice derivative only acts to the right. As a consequence, it is in general not a real but a complex number. The variational problem is usually conceived for a real action. Yet, as we show in Appendix \ref{app:symmetric_action}, this asymmetric action $S_{\text{DQW}}^{\text{asym.}} $ is equal (up to boundary terms) to the following \emph{symmetric} and therefore real-valued action,
\begin{equation}
\label{eq:action_symmetric}
S_{\text{DQW}} \defeq \epsilon \frac{\mathrm{i}}{2} \sum_n \bar{\bar{\psi}}_n  \tilde{\gamma}^{\mu} \psi_{n+\hat{\mu}} + \text{H.\ \hspace{-2mm} c.}  - \epsilon \sum_n \epsilon {m} \bar{\bar{\psi}}_n  {\psi}_n \, .
\end{equation}
One can show that extremalizing the symmetric action in Eq.\ \eqref{eq:action_symmetric} yields Eq.\ \eqref{eq:DQW_fermions}: this is proven in a more general case in Appendix \ref{app:current_conservation}, where we obtain the Euler-Lagrange equations from a generic real action.
That being said, note that boundary terms need not be taken into account in variational problems that determine the equations of motion\footnote{This is also true in the continuum. However, boundary terms \emph{do} need to be taken into account in variational problems of the type of Noether's theorem, see Sec.\ \ref{sec:Noether} and Appendix \ref{app:Noether} for present discrete setting.}; hence, since the symmetric and the asymmetric actions only differ by boundary terms, a proof with one of the two (for the type of variational problem mentioned) yields the corresponding result for the other one.

\section{Noether's theorem in discrete spacetime for internal symmetries}
\label{sec:Noether}

In this section, we derive a lattice Noether's theorem for internal symmetries. By ``internal symmetry'' we mean, as usual, that the corresponding transformation acts only on the internal Hilbert space of the system, and does not affect the spacetime coordinates. 

\subsection{The framework}
\label{subsec:framework}

Consider an action being the sum, over the lattice sites, of a Lagrangian density which is function of (i) the fields $\psi^a$, $a=1,2$, and (ii) their shifts in time and space, that is,
\begin{equation}
\label{eq:Lagrangian_density}
S_F \defeq \sum_n \mathscr{L}(\psi_n,\psi_{n+\hat{\mu}}, \psi_n^{\dag}, \psi^{\dag}_{n+\hat{\mu}}) \, ,
\end{equation}
where ``$F$'' stands for ``fermionic'' and where we consider, for the sake of correctness, a symmetric and therefore real-valued action\footnote{That being said, it turns out that in the concrete case $S_F = S_{\text{DQW}}$ all the computations that we are going to carry out can be carried out with $S_{\text{DQW}}^{\text{asym.}}$ instead of $S_{\text{DQW}}$ and still deliver the same results, at least in the current case of the U(1)-charge current. This observation is also a feature of the continuum theory. Notice that when two actions which differ by boundary terms are real-valued, the Noether currents are in general different.}.
Let us be fully precise on our notations:
here and below, $\phi_n=\psi_n,\psi_n^\dagger$ subsumes the family $(\phi^a_n)_a$ and, analogously, $\phi^a_{n+\hat{\mu}}$ subsumes the family $(\phi^a_{n+\hat{\mu}})_{\mu}$. 

The lattice Noether's theorem that we are going to derive is inspired by the general one for usual continuum field theories in Ref.\ \cite{SB05_notes}.
Consider an arbitrary transformation, $\psi_n^a \longrightarrow (\psi_n^a)' \defeq f^a((\psi_n^b)_b,\alpha)\equiv f^a(\psi_n,\alpha)$, of the field $\psi$, acting solely on its internal Hilbert space, where $\alpha$ denotes a family of real parameters\footnote{To consider transformations acting also on the external Hilbert space, one would have to supplement the transformation in Eq.\ \eqref{eq:transfo} by some coordinate transformation, see Ref.\ \cite{SB05_notes}.}.
In the present work, we will only consider the case of a global U(1) transformation, for which $\alpha$ reduces to a single real parameter. 
The generalization to a larger family of $\alpha$'s poses no major difficulty, but it renders the derivation more cumbersome and blurs its important aspects. 
Although the single application we know for $\alpha$ being a single real parameter is that of a global U(1) symmetry, we still present the proof for a general $f^a(\psi_n,\alpha)$ since this renders its generalization to a larger family of $\alpha$'s easier. 
The transformed state is collectively given by
\begin{equation}
\label{eq:transfo}
\psi'_n \defeq f(\psi_n,\alpha) \, .
\end{equation}
We parametrize the transformation $f$ such that
\begin{equation}
\label{eq:condition}
f(\psi_n,\alpha=0) = \psi_n \, .
\end{equation}
Finally, we assume $f$ to be differentiable and expand it to first order in the small parameter $\delta \alpha$ as
{\small
\begin{equation}
\psi'_n = f(\psi_n, \delta \alpha) = f(\psi_n,0) + \left.\frac{\partial f}{\partial \alpha}\right|_{(\psi_n,0)} \delta \alpha + O(\delta\alpha^2) \, .
\end{equation}}
Taking into account Eq.\ \eqref{eq:condition} and omitting higher-order terms this gives
\begin{equation}
\label{eq:field_transfo}
\psi'_n = \psi_n + C_n \delta \alpha \, ,
\end{equation}
where
\begin{equation}
C_n \defeq \left.\frac{\partial f}{\partial \alpha}\right|_{(\psi_n,0)} \, .
\end{equation}
In Appendix \ref{app:Euler-Lagrange}, we show how extremalizing the action in Eq.\ \eqref{eq:Lagrangian_density} yields an Euler-Lagrange equation for $\psi$.

\subsection{The Noether theorem}

The precise statement of our lattice Noether's theorem for internal symmetries is the following: if the generic action $S_F$ of Eq.\ \eqref{eq:Lagrangian_density} is invariant under the transformation in Eq.\ \eqref{eq:field_transfo}, i.e., if  Eq.\ \eqref{eq:field_transfo} is a(n) (internal) symmetry of the action, then the Noether current $J$ associated to the internal symmetry, and defined component-wise at site $n$ by
\begin{equation}
\label{eq:Kmu}
J^{\mu}_n \defeq \left. \frac{\partial\mathscr{L}}{\partial \psi_{n + \hat{\mu}}}\right|_{n} C_{n+\hat{\mu}} + C_{n+\hat{\mu}}^{\dag} \left.\frac{\partial\mathscr{L}}{\partial \psi^{\dag}_{n + \hat{\mu}}}\right|_{n} \, ,
\end{equation}
$\mu=0,1$, is conserved on the lattice (for a field $\psi$ that satisfies the Euler-Lagrange equation). That is, the lattice (one-step) $(1+1)$-divergence of $J$ vanishes, i.e.,
\begin{equation}
\label{eq:1plus1divergence}
d_{\mu}^L J^{\mu} = 0 \, ,
\end{equation}
where $d_\mu^L$ is the left lattice derivative defined in Eq.\ \eqref{eq:L-deriv}. 
We prove this theorem in Appendix \ref{app:Noether}.

\subsection*{Example: U(1) symmetry and charge conservation}
\label{subsec:U(1)}

Let us apply the general lattice Noether's theorem for internal symmetries of the preceding subsection, to the global U(1) symmetry of the (symmetric) DQW action, Eq.\ \eqref{eq:action_symmetric}.
The corresponding transformation is
\begin{equation}
f(\psi_n,\alpha) \defeq e^{\mathrm{i}\alpha}\psi_n \, ,
\end{equation}
so that the $C_n$ in Eq.\ \eqref{eq:field_transfo} is
\begin{equation}
C_n = \mathrm{i} \psi_n \, .
\end{equation}

The computation of the Noether current of Eq.\ \eqref{eq:Kmu}, which we call in the present case U(1)-charge current, delivers here, denoting it by $-J_{\text{U(1)}}$\footnote{The minus sign is to match with the most-used notations in the continuum limit.},
\begin{equation}
\label{eq:Jmu}
(J^{\mu}_{\text{U(1)}})_n \defeq \frac{\epsilon}{2} \left( \bar{\bar{\psi}}_{n} \tilde{\gamma}^{\mu} \psi_{n+\hat{\mu}} + \bar{\bar{\psi}}_{n+\hat{\mu}} \tilde{\gamma}^{\mu} \psi_{n}\right) \, ,
\end{equation}
where the $\tilde{\gamma}^{\mu}$'s have been defined in Eqs.\ \eqref{eqs:tildegammas}, and $\bar{\bar{\psi}}_n$  in Eq.\ \eqref{eq:psibar}.
The continuum limit of $(J^{\mu}_{\text{U(1)}})_n$ is trivially the well-known Dirac charge current (divide Eq.\ \eqref{eq:Jmu} by $\epsilon$ and let $\epsilon \rightarrow 0$), 
\begin{equation}
J^{\mu}_{\text{Dirac}}(x) \defeq  \bar{\psi}(x) \gamma^{\mu} \psi(x) \, ,
\end{equation}
where, as usual, $\bar{\psi} \defeq \psi^{\dag} \gamma^0$.

In Appendix \ref{app:current_conservation} we show how the lattice conservation equation satisfied by the U(1)-charge current in virtue of our lattice Noether's theorem, namely,
\begin{equation}
\label{eq:one-step_current-conservation}
 d^L_{\mu} J^{\mu}_{\text{U(1)}} = 0 \, ,  
\end{equation}
can be obtained, as in the continuum, from the EOM, either the one-step or directly the two-step.

\section{The $\mathrm{U}(1)$-gauged matter-field action and equations of motion}
\label{sec:gauging}

In this section, we are going to modify the matter-field action, Eq.\ \eqref{eq:action_symmetric}, in order to account for a coupling of the matter field $\psi$ to an Abelian U(1) lattice gauge field $A$ which is the gauge field of the continuum evaluated on the lattice. This gauge field is external for now, but it will become dynamical in the next section.

\subsection{The gauging procedure}

The usual gauging procedure for lattice systems is well-known for LGTs \cite{book_Rothe} and for DQWs\footnote{Ref.\ \cite{MMAMP18} achieved part of what is in Ref.\ \cite{CGWW18} regarding the gauging, but missed the proper gauging of the temporal translation operator.} \cite{CGWW18}: it consists in performing the following lattice \emph{minimal-coupling substitutions},
\begin{equation}
\label{eq:transformation}
\mathcal{T}_{\mu} \longrightarrow \mathcal{T}'_{\mu} \defeq  \mathcal{T}_{\mu}e^{-\mathrm{i}q\epsilon  A_{\mu}} \, ,
\end{equation}
where $q$ is the charge of the matter field $\psi$, and $A \defeq (A_{\mu})_{\mu=0,1}$ is the spacetime-dependent U(1) gauge field that we couple $\psi$ to. 
Applying the modified translation operators $\mathcal{T}'_{\mu}$ to the matter field $\psi$ at $n$ yields
\begin{equation}\label{eq:trafo_appl}
(\mathcal{T}_{\mu}e^{-\mathrm{i}q\epsilon A_{\mu}} \psi)_n = e^{-\mathrm{i}q\epsilon (A_{\mu})_{n-\hat{\mu}}} \psi_{n-\hat{\mu}} \, .
\end{equation}
The inverse of the modified translation operators,
\begin{equation}
\label{eq:transformation_inverse}
(\mathcal{T}'_{\mu})^{\dag} = e^{\mathrm{i}q\epsilon A_{\mu}} \mathcal{T}_{\mu}^{-1} \, ,
\end{equation}
accordingly act as
\begin{equation}
\label{eq:hopping}
(e^{\mathrm{i}q\epsilon A_{\mu}} \mathcal{T}_{\mu}^{-1}  \psi)_n = e^{\mathrm{i}q\epsilon (A_{\mu})_{n}} \psi_{n+\hat{\mu}} \, .
\end{equation}

\subsection{The gauged action and two-step equation of motion}
\label{subsec:gauging_two-step_scheme}

To couple $\psi$ to the gauge field $A_\mu$, we first rewrite the  matter-field action, Eq.\ \eqref{eq:action_symmetric}, as
\begin{equation}
\label{eq:action_reformulated}
S_{\text{DQW}} = \epsilon \frac{\mathrm{i}}{2} \sum_n \bar{\bar{\psi}}_n  \tilde{\gamma}^{\mu} (\mathcal{T}^{-1}_{\mu} \psi)_{n} + \text{H.c.}  - \epsilon \sum_n \epsilon {m}   \bar{\psi}_n  {\psi}_n \, .
\end{equation}
Performing the substitution \eqref{eq:transformation_inverse} (and \eqref{eq:transformation} in the Hermitian-conjugate term) on this action, we end up with the ``gauged'' action,
\begin{equation}
\label{eq:action_gauged}
S_{\text{DQW}}^{\text{g}} = \epsilon \frac{\mathrm{i}}{2} \sum_n \bar{\bar{\psi}}_n  \tilde{\gamma}^{\mu} e^{\mathrm{i}q\epsilon (A_{\mu})_n} \psi_{n+\hat{\mu}} + \text{H.c.}  - \epsilon \sum_n \epsilon {m}  \bar{\bar{\psi}}_n  {\psi}_n \, .
\end{equation}
This gauged action is exactly that of standard LGT, see Ref.\ \cite[Chapter 5]{book_Rothe}, up to substituting $\gamma^{\mu} \longrightarrow \tilde{\gamma}^{\mu}$ and $\bar{\psi} \longrightarrow \bar{\bar{\psi}}$.
It is invariant under the following gauge transformation,
\begin{subequations}
\label{eq:gauge_transfo}
\begin{align}
\psi_n &\longrightarrow \psi'_n \defeq G_n\psi_n \defeq e^{\mathrm{i}q\varphi_n}\psi_n \label{eq:gauge_transfo_psi}\\
(A_{\mu})_n &\longrightarrow  (A'_{\mu})_n \defeq (A_{\mu})_n - d^{R}_{\mu}\varphi|_n \, ,
\label{eq:gauge_transfo_field}
\end{align} 
\end{subequations}
where $d^R_\mu$ is the right lattice derivative defined in Eq.\ \eqref{eq:R-deriv}, and $\varphi_n$ is an arbitrary spacetime-dependent field.

Writing the Euler-Lagrange equation, 
\begin{equation}
    \left. \frac{\partial\mathscr{L}}{\partial \psi^{\dag}_n} \right|_n +  \left. \frac{\partial\mathscr{L}}{\partial \psi^{\dag}_{n+\hat{\mu}}} \right|_{n-\hat{\mu}}=0 \, ,
\end{equation}
derived in Appendix \ref{app:Euler-Lagrange}, for the action $S_{F} = S^{\text{g}}_{\text{DQW}}$, results in the following two-step EOM which we call U(1)-\emph{gauged unitary  fermions},
\begin{equation}
\label{eq:gauged_unitary__fermions}
\tilde{\gamma}^{\mu} \frac{\mathrm{i}}{2}\left( e^{\mathrm{i}q\epsilon (A_{\mu})_n} \psi_{n+\hat{\mu}} - e^{-\mathrm{i}q\epsilon (A_{\mu})_{n-\hat{\mu}}}\psi_{n-\hat{\mu}}  \right) - \epsilon {m} \psi_n = 0 \, ,
\end{equation}
which can  easily be shown to be exactly the EOM that we obtain if we directly gauge the EOM of unitary fermions, Eq.\ \eqref{eq:DQW_fermions}. 
This EOM is invariant under the Abelian U(1) gauge transformation \eqref{eq:gauge_transfo}.

\subsection{About gauging the one-step equation of motion}

In Eq.\ \eqref{eq:action_symmetric} we have considered an action that delivers the \emph{two-step} scheme, Eq.\ \eqref{eq:DQW_fermions}.
When coupling the matter field to a gauge field, it is therefore natural to apply the gauging procedure \eqref{eq:transformation} and \eqref{eq:transformation_inverse} to the two-step scheme, as done in the previous subsection. 
Yet, the unitarity of our model relies on the fact that -- without gauge fields -- we can find a unitary one-step scheme, Eq.\ \eqref{eq:sum-up}, that generates the two-step scheme.  
This immediately leads to the question  whether there is any unitary one-step scheme that generates the \emph{gauged} two-step scheme?
A natural follow-up question is then:
is the gauged two-step scheme \eqref{eq:gauged_unitary__fermions} generated by the gauged version of the one-step scheme? Equivalently, do the operations ``generating a two-step scheme'' and ``gauging'' commute? 
We are going to answer both questions in the affirmative, but only under a certain condition on the gauge field.

\subsubsection{Gauging the unitary one-step scheme}

The gauging of one-step discrete-spacetime systems is described in Ref.\ \cite{CGWW18}, and is in essence the same as that known in LGT: it consists in substituting translation operators as in \eqref{eq:transformation} and \eqref{eq:transformation_inverse}. Thus, gauging the one-step EOM of Eq.\ \eqref{eq:one-step scheme} gives
\begin{equation}
\label{eq:gauge_one-step_scheme}
e^{\mathrm{i}q\epsilon (A_0)_j} \psi_{j+1} = (\mathcal{W}_{\text{g}})_j \psi_j \, ,
\end{equation}
where $(\mathcal W_{\text{g}})_j$ is the gauged walk operator of Eq.\ \eqref{eq:walk_operator} evaluated at time $j$, i.e.,
\begin{equation}
\label{eq:Ug}
(\mathcal{W}_{\text{g}})_j \defeq W_{-1} e^{\mathrm{i}q\epsilon (A_1)_j} \mathcal{T}^{-1}_1 + W_1 \mathcal{T}_1 e^{-\mathrm{i}q\epsilon (A_1)_j} + W_0\mathbbm{1} \, .
\end{equation}
Moreover, the appearance of the gauge field on the left-hand side of Eq.\ \eqref{eq:gauge_one-step_scheme} stems from gauging $\mathcal T_0^{-1}$ in Eq.\ \eqref{eq:one-step scheme}.
One can verify that the unitarity conditions on $e^{-\mathrm{i}q\epsilon (A_0)_j}(\mathcal{W}_{\text{g}})_j$ are the same as those for the ungauged scheme (see Ref.\ \cite[Appendix A]{Arnault2022}).

\subsubsection{Recovering the gauged two-step scheme}

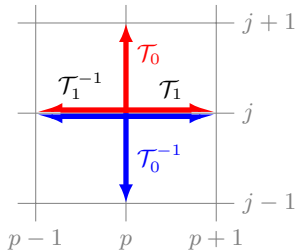
\begin{figure}[ht]
\begin{center}
    \begin{tikzpicture}[scale=1.2]
        every node/.append style={font=\footnotesize}
        \foreach \x/\y in {-1/-1,0/0,+1/+1}{
        \draw[help lines] (-1.2,\y) -- (1.2,\y) node[right] {\if\y0$j$\else$j\y$\fi}
                (\x,-1.2) node[below=9pt,anchor=base] {\if\x0$p$\else$p\x$\fi} -- (\x,1.2);
                        }
        \draw[line width=2pt,red,-{Latex[scale=.6]}] (0,0) -- node[right,pos=.65] {$\mathcal T_0$} (0,1); 
        \draw[line width=2pt,red,{Latex[right,scale=.6]}-{Latex[left,scale=.6]}] (-1,1pt) -- node[black,above] {$\mathcal T_1^{-1}$} (0,1pt) -- node[black,above] {$\mathcal T_1$} (1,1pt);
        \draw[line width=2pt,blue,-{Latex[scale=.6]}] (0,0) --  node[right] {$\mathcal T_0^{-1}$} (0,-1);
        \draw[line width=2pt,blue,{Latex[left,scale=.6]}-{Latex[right,scale=.6]}] (-1,-1pt) -- (1,-1pt);
        
    \end{tikzpicture}
\end{center}
\caption{\label{fig:alg_rels}The different links involved in Eq.\ \eqref{eq:gauge_one-step_scheme} (red) and Eq.\ \eqref{eq:gauged_second} (blue). Since $\mathcal T_0'$ and $\mathcal T_1'$ do not commute, the algebraic relations between the red and the blue links are different. All links, blue and red, are involved in the two-step scheme.}
\end{figure}

Multiplying Eq.\ \eqref{eq:gauge_one-step_scheme} on the left by $(\mathcal{W}_{\text{g}}^{\dag})_j\equiv ((\mathcal{W}_{\text{g}})_j)^{\dag}$ and shifting indices as $j\longrightarrow j-1$, we obtain
\begin{equation}
\label{eq:temporary}
    \psi_{j-1} = (\mathcal{W}^{\dag}_{\text{g}})_{j-1} e^{\mathrm{i}q\epsilon (A_0)_{j-1}}  \psi_j \, .
\end{equation}
Now, multiplying on the left by $e^{-\mathrm{i}q\epsilon (A_0)_{j-1}}$, we obtain
\begin{equation}
\label{eq:gauged_second}
e^{-\mathrm{i}q\epsilon (A_0)_{j-1}} \psi_{j-1} =  e^{-\mathrm{i}q\epsilon (A_0)_{j-1}}  (\mathcal{W}^{\dag}_{\text{g}})_{j-1} e^{\mathrm{i}q\epsilon (A_0)_{j-1}} \psi_j \, .  
\end{equation}
Note that this is different from simply evolving $\psi$ backwards in time by the inverse of $\mathcal W_{\text{g}}$, i.e., in general $e^{-\mathrm{i}q\epsilon (A_0)_{j-1}}  (\mathcal{W}^{\dag}_{\text{g}})_{j-1} e^{\mathrm{i}q\epsilon (A_0)_{j-1}} \neq (\mathcal{W}^{\dag}_{\text{g}})_{j-1}$. This is due to the presence of the gauge fields, i.e., the fact that $\mathcal T_0'$ and $\mathcal T_1'$ do not commute, see also Fig. \ref{fig:alg_rels}.

Substracting Eq.\ \eqref{eq:gauged_second} from Eq.\ \eqref{eq:gauge_one-step_scheme} and multiplying by $\mathrm{i}/2$ yields the EOM
\begin{align}
\label{eq:summ}
&\frac{\mathrm{i}}{2} \left( e^{\mathrm{i}q\epsilon (A_0)_j} \psi_{j+1}  - e^{-\mathrm{i}q\epsilon (A_0)_{j-1}} \psi_{j-1} \right) \\
& \ \ \ \ \ \ \ \ = \frac{\mathrm{i}}{2} \left( (\mathcal{W}_{\text{g}})_j - e^{-\mathrm{i}q\epsilon (A_0)_{j-1}}  (\mathcal{W}^{\dag}_{\text{g}})_{j-1} e^{\mathrm{i}q\epsilon (A_0)_{j-1}} \right) \psi_j \, , \nonumber
\end{align}
which is invariant under the $\text{U}(1)$ gauge transformation \eqref{eq:gauge_transfo}.

The left-hand side of Eq.\ \eqref{eq:summ} is indeed the gauged version of the left-hand side of the (generic) two-step scheme, Eq.\ \eqref{eq:generic_two-step_scheme}. However, the right-hand side is \emph{not} the gauged version of the lattice Hamiltonian operator defined directly at the two-step level, i.e., $(\mathcal{H}_{\text{g}})_j\defeq\frac{\mathrm{i}}{2}[(\mathcal{W}_{\text{g}})_j - ((\mathcal{W}^{\dag})_{\text{g}})_{j-1}]$, because in general $e^{-\mathrm{i}q\epsilon (A_0)_{j-1}}  (\mathcal{W}^{\dag}_{\text{g}})_{j-1} e^{\mathrm{i}q\epsilon (A_0)_{j-1}} \neq ((\mathcal{W}^{\dag})_{\text{g}})_{j-1}$.
So, the lesson to take away is that applying the gauging procedure to a one-step scheme and then deducing from it a two-step scheme, Eq.\ \eqref{eq:summ}, does not lead to the same two-step scheme as that obtained from applying the gauging procedure directly to the two-step scheme, Eq.\ \eqref{eq:gauged_unitary__fermions}.

What we can say is that $((\mathcal{W}^{\dag})_{\text{g}})_j = (\mathcal{W}^{\dag}_{\text{g}})_j$, which holds even in the non-Abelian case (this is easy to show). 
Given the previous identity, a sufficient condition for the gauging of the one-step scheme to produce as a two-step scheme the directly gauged version of the two-step scheme, i.e., a sufficient condition to have $e^{-\mathrm{i}q\epsilon (A_0)_{j-1}}  (\mathcal{W}^{\dag}_{\text{g}})_{j-1} e^{\mathrm{i}q\epsilon (A_0)_{j-1}} = (\mathcal{W}^{\dag}_{\text{g}})_{j-1}$, is that $(A_0)_n$ be independent of the spatial position; one may also take the well-known, stronger condition $A_0=0$, known as the temporal gauge.

\section{The action of the gauge field}
\label{sec:action_gauge_field}

In this section, we consider a $(3+1)$-dimensional spacetime unless otherwise mentioned.

\subsection{Link variables}

In Sec.\ \ref{sec:gauging} above, we introduced a lattice gauge field $(A_{\mu})_n$ that appears only in the form of the exponential
\begin{equation}\label{eq:link_var}
(U_{\mu})_n \defeq e^{\mathrm{i}q\epsilon (A_{\mu})_n} \, .
\end{equation}
As in the continuum, the gauge field $(A_{\mu})_n$ depends on a lattice site $n$ and a direction $\mu$.
On the lattice, this can be rephrased as a dependence on two neighbouring lattice sites $n$ and $n+\hat\mu$.
As in the LGT literature, we thereore call $(U_{\mu})_n$ a \emph{link variable}.
This link variable is conventionally associated with a hopping from $n$ to $n+\hat \mu$, and we stick to this convention here. Accordingly, we denote the link variable by $U_{n,n+\hat\mu}$ \cite{book_Rothe}, and its inverse by $U_{n+\hat\mu,n}:=(U_{\mu}^\dagger)_n$.

\subsection{Lattice field strength}

The gauge transformation of the gauge field, Eq.\ \eqref{eq:gauge_transfo_field}, reads, at the level of the link variable,
\begin{equation}
\label{eq:law_gauge}
    (U_{\mu})_n \longrightarrow (U_{\mu}')_n \defeq G_n  (U_{\mu})_n  G^{-1}_{n+\hat{\mu}} \, ,
\end{equation}
where $G_n \defeq e^{\mathrm{i}q\varphi_n}$ as in the transformation \eqref{eq:gauge_transfo_psi}. 
The gauge transformation of the adjoint link variable is
\begin{equation}
\label{eq:law_gauge_adjoint}
    (U_{\mu}^\dagger)_n \longrightarrow ({U_{\mu}^\dagger}')_n \defeq G_{n+\hat{\mu}} (U_{\mu}^\dagger)_n  G_n^{-1} \, .
\end{equation}

In the continuum theory, the gauge-field action, that determines the dynamics of the gauge field, is constructed from the so-called field strength $F_{\mu\nu}$. In the Abelian case, $F_{\mu\nu}$ is given by
\begin{equation}
F_{\mu\nu} \defeq \partial_{\mu} A_{\nu} - \partial_{\nu} A_{\mu}  \, .
\end{equation}
It is gauge invariant and determines the continuum gauge-field action $S_G^{\text{cont.}}$ via
\begin{equation}
S_G^{\text{cont.}} \defeq - \frac{1}{4} \int d^4x F_{\mu \nu}(x) F^{\mu\nu}(x) \, .
\end{equation}

\begin{figure}[ht]
\begin{center}
    \begin{tikzpicture}[scale=1.8]
        \foreach \x/\y in {0/0,1/1}{
        \draw[help lines] (-.2,\y) -- (1.2,\y)
                        (\x,-.2) -- (\x,1.2);
                        }
        \draw[very thick,red,-latex] (0,0) -- node[left] {$(U_0)_n$} (0,1);
        \draw[very thick,red,-latex] (1,0) -- node[right] {$(U_0)_{n+\hat1}$} (1,1) ;
        \draw[very thick,blue,-latex] (0,0) -- node[below] {$(U_1)_n$} (1,0);
        \draw[very thick,blue,-latex] (0,1) -- node[above] {$(U_1)_{n+\hat0}$} (1,1) ;
    \end{tikzpicture}
\end{center}
\caption{\label{fig:plaque}Terms involved in the plaquette operator $U_{10}$ from Eq.\ \eqref{eq:plaq_op}.}
\end{figure}
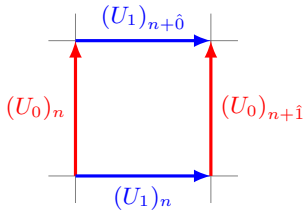

To build an action for the gauge field in discrete spacetime, we follow standard LGT and also Ref.\ \cite{CGWW18}. Multiplying the link variables along the smallest possible path on the lattice yields the so-called \emph{plaquette} operator,
\begin{equation}
(U_{\mu\nu})_n \defeq (U_{\mu})_n (U_{\nu})_{n+\hat{\mu}} (U_{\mu}^{\dag})_{n+\hat{\nu}} (U_{\nu}^{\dag})_n \, .   \label{eq:plaq_op}
\end{equation}
When viewing the link variable as a directed quantity the plaquette operator reads $(U_{\mu\nu})_n = U_{n,n+\hat{\mu}}U_{n+\hat{\mu},n+\hat{\mu}+\hat{\nu}} U_{n+\hat{\mu}+\hat{\nu}, n+\hat{\nu}} U_{n+\hat{\nu},n}$, where the indices follow a path which is an elementary square in the $\mu\nu$ plane on the lattice, see Fig.\ \ref{fig:plaque}.

This plaquette operator is in general gauge covariant, i.e., its transformation law under a change of gauge is
\begin{equation}
 (U_{\mu\nu})_n \longrightarrow  (U_{\mu\nu}')_n = G_n (U_{\mu\nu})_n   G_n^{-1} \, .
\end{equation}
In the Abelian case this implies that $U_{\mu\nu}$ is gauge invariant, i.e.,
\begin{equation}
(U_{\mu\nu}')_n = (U_{\mu\nu})_n  \, .
\end{equation}

Inserting the definition of the link variables, Eq.\ \eqref{eq:link_var}, into Eq.\ \eqref{eq:plaq_op}, we can express the Abelian plaquette operator also as
\begin{equation}
 (U_{\mu\nu})_n = e^{\mathrm{i}q\epsilon^2(F_{\mu\nu})_n} \, ,  
\end{equation}
where $(F_{\mu\nu})_n$ is the lattice field strength
\begin{equation}
(F_{\mu\nu})_n = d^R_{\mu} A_{\nu}|_n -  d^R_{\nu} A_{\mu}|_n \, ,   
\end{equation}
where we recall that $d^R_{\mu}$  is the right lattice derivative, defined in Eq.\ \eqref{eq:R-deriv}.

\subsection{Lattice gauge-field action}

We now have to construct from the lattice gauge-invariant quantity $(U_{\mu\nu})_n$ an action that has the correct continuum limit: we show in Appendix \ref{app:gauge-field_action} that the lattice action
\begin{equation}
\label{eq:gauge-field_action}
    S_G \defeq S_G^{\text{time}} + S_G^{\text{space}} \, ,
\end{equation}
where
\begin{subequations}
\label{eqs:actions}
\begin{align}
    S_G^{\text{time}} &\defeq \frac{1}{q^2} \sum_n \sum_l \left[1 -  \frac{1}{2}\left[ (U_{0l})_n + (U_{0l})^{\dag}_n\right]  \right] \\
     S_G^{\text{space}} &\defeq \frac{1}{q^2} \sum_n \sum_{\substack{k,l \\ k<l}} \left[\frac{1}{2}\left[(U_{kl})_n + (U_{kl})^{\dag}_n\right] - 1 \right] \, ,
\end{align}
\end{subequations}
has the correct continuum limit $S^{\text{cont.}}_G$ \cite{Weinberg_QFT1}, which can be seen from expanding $S_G$ in $\epsilon$ and rewriting it as
\begin{equation}
\label{eq:cont_limit}
    S_G = -\frac{1}{4} \sum_n \epsilon^4  (F_{\mu \nu})_n (F^{\mu\nu})_n + O(\epsilon^6) \, .
\end{equation}
In $1+1$ dimensions, $S_G^{\text{space}}$ trivially vanishes. 
Note that $S_G^{\text{time}}$ is a purely electric term, i.e., it involves only the lattice electric field \cite{ced13,CW19} with components $(F_{0l})_n$, while $S_G^{\text{space}}$ is a purely magnetic term, i.e., it involves only the lattice magnetic field with components $(F_{kl})_n$.
Note that the suggested action, Eq.\ \eqref{eq:gauge-field_action}, is nothing but a real-time version of the Euclidean one in Eq.\ (5.21) of Ref.\ \cite{book_Rothe}.

\subsection{The classical lattice dynamics of the gauge field}

In this subsection, we consider $S_G$ in a $(1+1)$-dimensional spacetime, so that in Eqs.\ \eqref{eqs:actions} one must replace $1/q^2$ by $1/(q^2\epsilon^2)$.
Still, we continue to write the magnetic terms in Secs.\ \ref{eq:subsubsec:euler-lag} and \ref{eq:subsubsec:inhomogeneous} as if we were in $3+1$ dimensions, in order to make the equations look more general.
For these equations to actually make sense in more than $1+1$ dimensions one would have to find a $(3+1)$-dimensional version of $S^{\text{g}}_{\text{DQW}}$, which would give sense to a $(3+1)$-dimensional U(1)-charge current $(J^{\text{g}}_{\text{U(1)}})^{\mu}_n$  below; 
this has not been done in the present paper. 
Alternatively, one can assume to have found a $(3+1)$-dimensional version of $S^{\text{g}}_{\text{DQW}}$, which gives sense to a $(3+1)$-dimensional U(1)-charge current $(J^{\text{g}}_{\text{U(1)}})^{\mu}_n$ below, and then consider a $(3+1)$-dimensional version of $S_G$;
note however that this leads to additional $\epsilon$ factors in the right-hand sides of Eqs.\ \eqref{eq:Jmugauged} and \eqref{eq:right_term}.

\subsubsection{Lattice Euler-Lagrange equations for the gauge field}

Consider the total action
\begin{equation}
S \defeq S_{\text{DQW}}^{\text{g}} + S_G \, ,
\end{equation}
where $S_{\text{DQW}}^{\text{g}}$ is given in Eq.\ \eqref{eq:action_gauged} and $S_G$ in Eq.\ \eqref{eq:gauge-field_action}.
We may be tempted to consider $S$ as a function of the link variables (and their translates) rather than of the $(A_{\mu})_n$'s (and their discrete derivatives), but then we would also have to take into account the adjoints of the link variables. If, instead, we take $S$ as a function of the $(A_{\mu})_n$'s no such question arises.
Hence, we consider $S$ as the following action functional,
\begin{equation}
\label{eq:total_action}
S = \sum_n \mathscr{L}\left( ((A_{\mu})_n)_{\mu=0,1}, (d^R_{\nu}A_{\mu}|_n)_{\mu,\nu=0,1} \right) \, .
\end{equation}
Extremalizing this action, Eq.\ \eqref{eq:total_action}, leads to the following Euler-Lagrange equations of motion for the gauge field $A_{\mu}$,
\begin{equation}
\label{eq:Euler-Lagrange_gauge_field}
   \left. \frac{\partial\mathscr{L}}{\partial (A_{\mu})_n} \right|_n - \left. d^L_{\nu} \frac{\partial \mathscr{L}}{\partial \, d^R_{\nu} A_{\mu}|_n} \right|_n = 0 \, ,
\end{equation}
where we recall that $d^L_\mu$ and $d^R_\mu$ are the left and right lattice derivatives, defined in Eqs. \eqref{eq:LR-derivs}.

\subsubsection{Evaluating the lattice Euler-Lagrange equations}
\label{eq:subsubsec:euler-lag}

The first term of Eq.\ \eqref{eq:Euler-Lagrange_gauge_field} is easily computed, and yields
\begin{equation}
\label{eq:left_term}
 \left. \frac{\partial\mathscr{L}}{\partial (A_{\mu})_n} \right|_n  = - q \epsilon (J^{\text{g}}_{\text{U(1)}})^{\mu}_{n} \, ,   
\end{equation}
where
\begin{equation}
\label{eq:Jmugauged}
\begin{split}
&(J^{\text{g}}_{\text{U(1)}})^{\mu}_{n} \\
&\defeq \frac{\epsilon}{2} \left( \bar{\bar{\psi}}_{n} \tilde{\gamma}^{\mu} e^{\mathrm{i}q\epsilon(A_{\mu})_n}\psi_{n+\hat{\mu}} + \bar{\bar{\psi}}_{n+\hat{\mu}} e^{-\mathrm{i}q\epsilon(A_{\mu})_n} \tilde{\gamma}^{\mu} \psi_{n}\right) \, ,
\end{split}
\end{equation}
which coincides with the gauged version of the Noether U(1)-charge current given in Eq.\ \eqref{eq:Jmu}.

To compute the second term of the Euler-Lagrange equations, we need to distinguish between $\mu=0$ and $\mu = l \neq 0$.
After a few lines of computation for $\mu=0$, remembering that $F_{\nu0}=-F^{\nu0}$, another few lines for $\nu=0$, and another few lines for $\nu=k$ and $\mu=l$, remembering that $F_{kl}=F^{kl}$, we combine all three formul{\ae} into
\begin{equation}
\label{eq:right_term}
- \left. d^L_{\nu} \frac{\partial \mathscr{L}}{\partial \, d^R_{\nu} A_{\mu}|_n} \right|_n =  d^L_{\nu} \left[\frac{1}{q} \sin \left( q\epsilon^2 F^{\nu\mu}_n \right) \right] \, .  
\end{equation}

Inserting Eqs.\ \eqref{eq:left_term} and \eqref{eq:right_term} into Eq.\ \eqref{eq:Euler-Lagrange_gauge_field}, we obtain the following equations of motion for the gauge field,
\begin{equation}
\label{eq:Maxwell}
d^L_{\nu} \left[ \frac{1}{q}  \sin \left( q\epsilon^2 F^{\nu\mu}_n \right) \right] = q \epsilon (J^{\text{g}}_{\text{U(1)}})^{\mu}_{n} \, .   
\end{equation}

\subsubsection{The two inhomogeneous lattice Maxwell equations}
\label{eq:subsubsec:inhomogeneous}

Let us show that the equations of motion for the gauge field, Eq.\ \eqref{eq:Maxwell}, correspond to the two inhomogeneous lattice Maxwell equations.
For $\mu=0$, Eq. \eqref{eq:Maxwell} gives
\begin{equation}
\label{eq:lattice_Maxwell-Gauss}
d^L_{k}  \left[ \frac{1}{q}  \sin \left( q \epsilon^2 E^k_n\right) \right] = q \epsilon (J^{\text{g}}_{\text{DQW}})^{0}_{n} \, , 
\end{equation}
where
\begin{equation}
E^k_n \defeq F^{k0}_n \, ,    
\end{equation}
is the lattice electric field.
Equation \eqref{eq:lattice_Maxwell-Gauss} is a lattice version of Maxwell-Gauss' equation, which is a constraint rather than a dynamical equation: dividing by $\epsilon^2$ on both sides and taking the limit $\epsilon \rightarrow 0$ indeed yields
\begin{equation}
    \partial_k E^k = q J^{0}_{\text{Dirac}} \, .
\end{equation}

For $\mu=l$, Eq. \eqref{eq:Maxwell} becomes
{\small
\begin{equation}
d^L_0 \left[ \frac{1}{q}  \sin \left(q\epsilon^2 F^{0l}_n \right)\right] +  d^L_k  \left[ \frac{1}{q} \sin \left(q\epsilon^2 F^{kl}_n \right) \right] = q \epsilon (J^{\text{g}}_{\text{DQW}})^{l}_{n} \, ,
\end{equation}}
that is, taking into account 
\begin{align}
    F^{kl}_n &\equiv - {\varepsilon^{kl}}_m B^m_n \, ,
\end{align}
where $B^m_n$ is the lattice magnetic field, and swapping $k$ and $l$ in ${\varepsilon^{kl}}_m$ (so that we pick up a minus sign),
{\small
\begin{equation}
\label{eq:Maxwell_ampere}
- d^L_0 \left[ \frac{1}{q}  \sin \left(q\epsilon^2 E^l_n \right) \right] + {\varepsilon^{lk}}_m d^L_k \left[  \frac{1}{q}  \sin \left(q\epsilon^2 B^{m}_n\right) \right] = q \epsilon (J^{\text{g}}_{\text{DQW}})^{l}_{n} \, .
\end{equation}}
\noindent
This is a lattice version of Maxwell-Ampère's equation, which can be seen by taking in Eq.\ \eqref{eq:Maxwell_ampere}  the limit $\epsilon \rightarrow 0$ after having divided by $\epsilon^2$ on both sides, which yields
\begin{equation}
  - \partial_0 E^l + {\varepsilon^{lk}}_m \partial_k B^m = q J_{\text{Dirac}}^l \, .
\end{equation}
The convergence of these two lattice Maxwell equations, Eqs.\ \eqref{eq:lattice_Maxwell-Gauss} and \eqref{eq:Maxwell_ampere}, has been proven in Ref.\ \cite{CH09}.

\subsubsection{The two inhomogeneous lattice Maxwell equations in $(1+1)$-dimensional spacetime}

In $(1+1)$-dimensional spacetime, the lattice Maxwell-Gauss constraint, Eq.\ \eqref{eq:lattice_Maxwell-Gauss}, reduces to
\begin{equation}
d^L_{1}  \left[ \frac{1}{q}  \sin \left( q \epsilon^2 E^1_n\right) \right] = q \epsilon (J^{\text{g}}_{\text{DQW}})^{0}_{n} \, ,
\end{equation}
with continuum limit
\begin{equation}
       \partial_1 E^1 = q J^{0}_{\text{Dirac}} \, ,
\end{equation}
and the lattice Maxwell-Ampère equation, Eq.\  \eqref{eq:Maxwell_ampere}, reduces to
\begin{equation}
\label{eq:Maxwell_ampere_11}
- d^L_0 \left[ \frac{1}{q}  \sin \left(q\epsilon^2 (E^1)_n \right) \right]  = q \epsilon  (J^{\text{g}}_{\text{DQW}})^{1}_{n} \, ,
\end{equation}
with continuum limit
\begin{equation}
  - \partial_0 E^1 = q J_{\text{Dirac}}^1 \, .
\end{equation}

Finally, the homogeneous Maxwell equations, namely, Maxwell-Thompson's and Maxwell-Faraday's equations, are not relevant in one spatial dimension.

\section{Conclusions and discussion}

Let us sum up the main achievements of this work.
First, we have constructed a discrete-spacetime action $S_{\text{DQW}}$ for a spin-1/2 matter field, with the following properties: it is (i) in real time, (ii) based on a two-step EOM and therefore extremely similar to usual actions of LGT, (iii) associated to a unitary (classical-fields) one-step scheme, i.e., a discrete-time quantum walk, which in the continuum limit yields the Dirac equation.
Second, we have derived a lattice Noether's theorem for internal symmetries of $S_{\text{DQW}}$.
More precisely, we have proven the theorem for an internal symmetry depending on a single real parameter $\alpha$, but believe that the generalization to a larger parameter family poses no major difficulty.
We have applied the Noether's theorem to the global $\text{U}(1)$ symmetry of $S_{\text{DQW}}$ and have obtained a conserved current which in the continuum coincides with the usual Dirac charge current.
Third, we have coupled $S_{\text{DQW}}$ via a minimal coupling on the lattice to an Abelian U(1) gauge field. 
Although $S_{\text{DQW}}$ is based on a two-step EOM, unitarity relies on the fact that this two-step EOM is associated to a one-step EOM. We have thus explored the gauging of the one-step EOM as in Ref.\ \cite{CGWW18}. This has lead us to the observation that gauging directly the two-step EOM does not yield the same EOM as that obtained by first gauging the one-step EOM and then constructing a two-step EOM from it; the two procedures are equivalent only if the temporal component of the gauge field, $A_0$, is independent of space.
Finally, we have suggested a real-time LGT-type action for the Abelian U(1) gauge field, from which we have derived the classical EOMs of the gauge field, which are lattice versions of Maxwell's equations.

A first question that remains unanswered is the following: Is it true that there exist no underlying unitary scheme for naive fermions? While we believe this to be the case, we do not have a proof.
A second topic that is unadressed in this work is that of external (i.e., spacetime) symmetries of the action:
Can one define such symmetries and derive associated Noether's theorems? Refs.\ \cite{AFF14a, Debbasch2019b} partially address this problem in the realm of DQWs.
Also, we did not address quantized fields:
What multi-particle concepts from the QCA LGT in Refs.\ \cite{ABF20, FS2020, SADM2022, EDMMplus22} should we import into our Lagrangian framework in order to build a fully fledged action-based LGT that respects strict locality and describes fermionic matter fields? 

Finally, the issue of fermion doubling has not been adressed, because it is still unsolved for the following reasons. In Ref.\ \cite{Arnault2022}  a fermion-doubling issue is solved for the two-step ``Hamiltonian''. 
But, although such a two-step Hamiltonian would indeed be subject to a fermion-doubling issue if the scheme was in continuous time, in Ref.\ \cite{Arnault2022} time is discrete. In addition, the two-step scheme is generated by a one-step scheme which does \emph{not} exhibit a fermion-doubling issue \cite{book_Rothe}. One should hence examine whether the fact that the two-step Hamiltonian of Ref.\ \cite{Arnault2022} has a fermion-doubling issue is actually relevant if time is discrete and, moreover, if the generating one-step scheme does not have this issue. Now, all that being said, in the present framework  the matter-field action $S_{\text{DQW}}$, that is defined with the two-step Hamiltonian, is extremely similar to usual LGT fermionic actions \cite{Hernandez2011}, which suggests that a fermion-doubling issue \emph{does} arise. To sum up: if $S_{\text{DQW}}$ does not exhibit fermion doubling, then there is nothing to be done, and if it does, then modifying the two-step Hamiltonian as in Ref.\ \cite{Arnault2022} should remove at least spatial doublers. Finally, if there are temporal doublers in $S_{\text{DQW}}$, then some fix must be made in order to remove them without breaking the unitarity of the model.

\section*{Acknowledgements}
C. Cedzich was supported in part by the Deutsche Forschungsgemeinschaft (DFG, German Research Foundation) under the grant number 441423094.
The authors thank Pablo Arrighi for his professional support.


\begin{thebibliography}{76}%
\makeatletter
\providecommand \@ifxundefined [1]{%
 \@ifx{#1\undefined}
}%
\providecommand \@ifnum [1]{%
 \ifnum #1\expandafter \@firstoftwo
 \else \expandafter \@secondoftwo
 \fi
}%
\providecommand \@ifx [1]{%
 \ifx #1\expandafter \@firstoftwo
 \else \expandafter \@secondoftwo
 \fi
}%
\providecommand \natexlab [1]{#1}%
\providecommand \enquote  [1]{``#1''}%
\providecommand \bibnamefont  [1]{#1}%
\providecommand \bibfnamefont [1]{#1}%
\providecommand \citenamefont [1]{#1}%
\providecommand \href@noop [0]{\@secondoftwo}%
\providecommand \href [0]{\begingroup \@sanitize@url \@href}%
\providecommand \@href[1]{\@@startlink{#1}\@@href}%
\providecommand \@@href[1]{\endgroup#1\@@endlink}%
\providecommand \@sanitize@url [0]{\catcode `\\12\catcode `\$12\catcode
  `\&12\catcode `\#12\catcode `\^12\catcode `\_12\catcode `\%12\relax}%
\providecommand \@@startlink[1]{}%
\providecommand \@@endlink[0]{}%
\providecommand \url  [0]{\begingroup\@sanitize@url \@url }%
\providecommand \@url [1]{\endgroup\@href {#1}{\urlprefix }}%
\providecommand \urlprefix  [0]{URL }%
\providecommand \Eprint [0]{\href }%
\providecommand \doibase [0]{https://doi.org/}%
\providecommand \selectlanguage [0]{\@gobble}%
\providecommand \bibinfo  [0]{\@secondoftwo}%
\providecommand \bibfield  [0]{\@secondoftwo}%
\providecommand \translation [1]{[#1]}%
\providecommand \BibitemOpen [0]{}%
\providecommand \bibitemStop [0]{}%
\providecommand \bibitemNoStop [0]{.\EOS\space}%
\providecommand \EOS [0]{\spacefactor3000\relax}%
\providecommand \BibitemShut  [1]{\csname bibitem#1\endcsname}%
\let\auto@bib@innerbib\@empty
\bibitem [{\citenamefont {Rothe}(2012)}]{book_Rothe}%
  \BibitemOpen
  \bibfield  {author} {\bibinfo {author} {\bibfnamefont {H.~J.}\ \bibnamefont
  {Rothe}},\ }\href {http://www.worldscientific.com/worldscibooks/10.1142/8229}
  {\emph {\bibinfo {title} {Lattice Gauge Theories: An Introduction{,} 4th
  {E}dition}}}\ (\bibinfo  {publisher} {World Scientific},\ \bibinfo {year}
  {2012})\BibitemShut {NoStop}%
\bibitem [{\citenamefont {Wilson}(1974)}]{Wilson74}%
  \BibitemOpen
  \bibfield  {author} {\bibinfo {author} {\bibfnamefont {K.~G.}\ \bibnamefont
  {Wilson}},\ }\bibfield  {title} {\bibinfo {title} {Confinement of quarks},\
  }\href {https://link.aps.org/doi/10.1103/PhysRevD.10.2445} {\bibfield
  {journal} {\bibinfo  {journal} {Phys. Rev. D}\ }\textbf {\bibinfo {volume}
  {10}},\ \bibinfo {pages} {2445} (\bibinfo {year} {1974})}\BibitemShut
  {NoStop}%
\bibitem [{\citenamefont {Hern{\'{a}}ndez}(2011)}]{Hernandez2011}%
  \BibitemOpen
  \bibfield  {author} {\bibinfo {author} {\bibfnamefont {M.~P.}\ \bibnamefont
  {Hern{\'{a}}ndez}},\ }\bibfield  {title} {\bibinfo {title} {Lattice field
  theory fundamentals},\ }in\ \href
  {https://doi.org/10.1093/acprof:oso/9780199691609.003.0001} {\emph {\bibinfo
  {booktitle} {Modern Perspectives in Lattice {QCD}: Quantum Field Theory and
  High Performance Computing, Lecture Notes of the Les Houches Summer School:
  Volume 93, August 2009}}}\ (\bibinfo  {publisher} {Oxford University Press},\
  \bibinfo {year} {2011})\ pp.\ \bibinfo {pages} {1--91}\BibitemShut {NoStop}%
\bibitem [{\citenamefont {Wilson}\ and\ \citenamefont {Kogut}(1974)}]{WK74}%
  \BibitemOpen
  \bibfield  {author} {\bibinfo {author} {\bibfnamefont {K.}~\bibnamefont
  {Wilson}}\ and\ \bibinfo {author} {\bibfnamefont {J.}~\bibnamefont {Kogut}},\
  }\bibfield  {title} {\bibinfo {title} {The renormalization group and the
  $\varepsilon$ expansion},\ }\href
  {https://doi.org/10.1016/0370-1573(74)90023-4} {\bibfield  {journal}
  {\bibinfo  {journal} {Phys. Rep.}\ }\textbf {\bibinfo {volume} {12}},\
  \bibinfo {pages} {75} (\bibinfo {year} {1974})}\BibitemShut {NoStop}%
\bibitem [{\citenamefont {Kogut}\ and\ \citenamefont
  {Susskind}(1975)}]{KogSuss75a}%
  \BibitemOpen
  \bibfield  {author} {\bibinfo {author} {\bibfnamefont {J.}~\bibnamefont
  {Kogut}}\ and\ \bibinfo {author} {\bibfnamefont {L.}~\bibnamefont
  {Susskind}},\ }\bibfield  {title} {\bibinfo {title} {Hamiltonian formulation
  of {W}ilson's lattice gauge theories},\ }\href
  {https://doi.org/10.1103/physrevd.11.395} {\bibfield  {journal} {\bibinfo
  {journal} {Phys. Rev. D}\ }\textbf {\bibinfo {volume} {11}},\ \bibinfo
  {pages} {395} (\bibinfo {year} {1975})}\BibitemShut {NoStop}%
\bibitem [{\citenamefont {Weinberg}(1995)}]{Weinberg_QFT1}%
  \BibitemOpen
  \bibfield  {author} {\bibinfo {author} {\bibfnamefont {S.}~\bibnamefont
  {Weinberg}},\ }\href
  {http://www.cambridge.org/catalogue/catalogue.asp?isbn=9780521670531} {\emph
  {\bibinfo {title} {The Quantum Theory of Fields I}}}\ (\bibinfo  {publisher}
  {Cambridge {U}niversity {P}ress},\ \bibinfo {year} {1995})\BibitemShut
  {NoStop}%
\bibitem [{\citenamefont {{M. C. Ba\~nuls}}\ \emph {et~al.}(2017)\citenamefont
  {{M. C. Ba\~nuls}}, \citenamefont {{K. Cichy}}, \citenamefont {{J. I.
  Cirac}}, \citenamefont {{K. Jansen}}, \citenamefont {{S. K\"uhn}},\ and\
  \citenamefont {{H. Saito}}}]{BCCJplus17}%
  \BibitemOpen
  \bibfield  {author} {\bibinfo {author} {\bibnamefont {{M. C. Ba\~nuls}}},
  \bibinfo {author} {\bibnamefont {{K. Cichy}}}, \bibinfo {author}
  {\bibnamefont {{J. I. Cirac}}}, \bibinfo {author} {\bibnamefont {{K.
  Jansen}}}, \bibinfo {author} {\bibnamefont {{S. K\"uhn}}},\ and\ \bibinfo
  {author} {\bibnamefont {{H. Saito}}},\ }\bibfield  {title} {\bibinfo {title}
  {Towards overcoming the {M}onte {C}arlo sign problem with tensor networks},\
  }\href {https://doi.org/10.1051/epjconf/201713704001} {\bibfield  {journal}
  {\bibinfo  {journal} {EPJ Web Conf.}\ }\textbf {\bibinfo {volume} {137}},\
  \bibinfo {pages} {04001} (\bibinfo {year} {2017})},\ \Eprint
  {https://arxiv.org/abs/1611.04791} {arXiv:1611.04791} \BibitemShut {NoStop}%
\bibitem [{\citenamefont {Lang}(2008)}]{Lang2008}%
  \BibitemOpen
  \bibfield  {author} {\bibinfo {author} {\bibfnamefont {C.~B.}\ \bibnamefont
  {Lang}},\ }\bibfield  {title} {\bibinfo {title} {The hadron spectrum from
  lattice {QCD}},\ }\href
  {https://doi.org/https://doi.org/10.1016/j.ppnp.2007.12.026} {\bibfield
  {journal} {\bibinfo  {journal} {Prog. Part. Nucl. Phys.}\ }\textbf {\bibinfo
  {volume} {61}},\ \bibinfo {pages} {35} (\bibinfo {year} {2008})},\ \Eprint
  {https://arxiv.org/abs/0711.3091} {arXiv:0711.3091} \BibitemShut {NoStop}%
\bibitem [{\citenamefont {{Aoki et al.}}(2021)}]{Aoki2021}%
  \BibitemOpen
  \bibfield  {author} {\bibinfo {author} {\bibfnamefont {Y.}~\bibnamefont
  {{Aoki et al.}}},\ }\bibfield  {title} {\bibinfo {title} {{FLAG} {R}eview
  2021},\ }\href {https://arxiv.org/abs/2111.09849} {\bibfield  {journal}
  {\bibinfo  {journal} {arXiv:2111.09849}\ } (\bibinfo {year}
  {2021})}\BibitemShut {NoStop}%
\bibitem [{\citenamefont {{R. Brower}}\ \emph {et~al.}(2018)\citenamefont {{R.
  Brower}}, \citenamefont {{N. Christ}}, \citenamefont {{C. De{T}ar}},
  \citenamefont {{R. Edwards}},\ and\ \citenamefont {{P.
  Mackenzie}}}]{BCDEplus2018}%
  \BibitemOpen
  \bibfield  {author} {\bibinfo {author} {\bibnamefont {{R. Brower}}}, \bibinfo
  {author} {\bibnamefont {{N. Christ}}}, \bibinfo {author} {\bibnamefont {{C.
  De{T}ar}}}, \bibinfo {author} {\bibnamefont {{R. Edwards}}},\ and\ \bibinfo
  {author} {\bibnamefont {{P. Mackenzie}}},\ }\bibfield  {title} {\bibinfo
  {title} {Lattice {QCD} application development within the {US DOE} exascale
  computing project},\ }\href {https://doi.org/10.1051/epjconf/201817509010}
  {\bibfield  {journal} {\bibinfo  {journal} {EPJ Web Conf.}\ }\textbf
  {\bibinfo {volume} {175}},\ \bibinfo {pages} {09010} (\bibinfo {year}
  {2018})},\ \Eprint {https://arxiv.org/abs/1710.11094} {arXiv:1710.11094}
  \BibitemShut {NoStop}%
\bibitem [{\citenamefont {Troyer}\ and\ \citenamefont {Wiese}(2005)}]{TW2005}%
  \BibitemOpen
  \bibfield  {author} {\bibinfo {author} {\bibfnamefont {M.}~\bibnamefont
  {Troyer}}\ and\ \bibinfo {author} {\bibfnamefont {U.-J.}\ \bibnamefont
  {Wiese}},\ }\bibfield  {title} {\bibinfo {title} {Computational complexity
  and fundamental limitations to fermionic quantum {M}onte {C}arlo
  simulations},\ }\href {https://doi.org/10.1103/PhysRevLett.94.170201}
  {\bibfield  {journal} {\bibinfo  {journal} {Phys. Rev. Lett.}\ }\textbf
  {\bibinfo {volume} {94}},\ \bibinfo {pages} {170201} (\bibinfo {year}
  {2005})},\ \Eprint {https://arxiv.org/abs/cond-mat/0408370}
  {arXiv:cond-mat/0408370} \BibitemShut {NoStop}%
\bibitem [{\citenamefont {Preskill}(2018)}]{Preskill2018}%
  \BibitemOpen
  \bibfield  {author} {\bibinfo {author} {\bibfnamefont {J.}~\bibnamefont
  {Preskill}},\ }\bibfield  {title} {\bibinfo {title} {{Simulating quantum
  field theory with a quantum computer}},\ }\href
  {https://doi.org/10.22323/1.334.0024} {\bibfield  {journal} {\bibinfo
  {journal} {Po{S}}\ }\textbf {\bibinfo {volume} {LATTICE2018}},\ \bibinfo
  {pages} {024} (\bibinfo {year} {2018})},\ \Eprint
  {https://arxiv.org/abs/1811.10085} {arXiv:1811.10085} \BibitemShut {NoStop}%
\bibitem [{\citenamefont {Feynman}(1982)}]{Feynman1982}%
  \BibitemOpen
  \bibfield  {author} {\bibinfo {author} {\bibfnamefont {R.~P.}\ \bibnamefont
  {Feynman}},\ }\bibfield  {title} {\bibinfo {title} {Simulating physics with
  computers},\ }\href {https://doi.org/10.1007/bf02650179} {\bibfield
  {journal} {\bibinfo  {journal} {Int. J. Theor. Phys.}\ }\textbf {\bibinfo
  {volume} {21}},\ \bibinfo {pages} {467} (\bibinfo {year} {1982})}\BibitemShut
  {NoStop}%
\bibitem [{\citenamefont {Nielsen}\ and\ \citenamefont
  {Chuang}(2010)}]{book_nielsen_chuang_2010}%
  \BibitemOpen
  \bibfield  {author} {\bibinfo {author} {\bibfnamefont {M.~A.}\ \bibnamefont
  {Nielsen}}\ and\ \bibinfo {author} {\bibfnamefont {I.~L.}\ \bibnamefont
  {Chuang}},\ }\href {https://doi.org/10.1017/CBO9780511976667} {\emph
  {\bibinfo {title} {Quantum Computation and Quantum Information: 10th
  Anniversary Edition}}}\ (\bibinfo  {publisher} {Cambridge University Press},\
  \bibinfo {year} {2010})\BibitemShut {NoStop}%
\bibitem [{\citenamefont {Zohar}\ and\ \citenamefont {Reznik}(2011)}]{ZR2011}%
  \BibitemOpen
  \bibfield  {author} {\bibinfo {author} {\bibfnamefont {E.}~\bibnamefont
  {Zohar}}\ and\ \bibinfo {author} {\bibfnamefont {B.}~\bibnamefont {Reznik}},\
  }\bibfield  {title} {\bibinfo {title} {Confinement and lattice
  quantum-electrodynamic electric flux tubes simulated with ultracold atoms},\
  }\href {https://doi.org/10.1103/PhysRevLett.107.275301} {\bibfield  {journal}
  {\bibinfo  {journal} {Phys. Rev. Lett.}\ }\textbf {\bibinfo {volume} {107}},\
  \bibinfo {pages} {275301} (\bibinfo {year} {2011})},\ \Eprint
  {https://arxiv.org/abs/1108.1562} {arXiv:1108.1562} \BibitemShut {NoStop}%
\bibitem [{\citenamefont {Banerjee}\ \emph {et~al.}(2012)\citenamefont
  {Banerjee}, \citenamefont {Dalmonte}, \citenamefont {M\"uller}, \citenamefont
  {Rico}, \citenamefont {Stebler}, \citenamefont {Wiese},\ and\ \citenamefont
  {Zoller}}]{BDMRplus2012}%
  \BibitemOpen
  \bibfield  {author} {\bibinfo {author} {\bibfnamefont {D.}~\bibnamefont
  {Banerjee}}, \bibinfo {author} {\bibfnamefont {M.}~\bibnamefont {Dalmonte}},
  \bibinfo {author} {\bibfnamefont {M.}~\bibnamefont {M\"uller}}, \bibinfo
  {author} {\bibfnamefont {E.}~\bibnamefont {Rico}}, \bibinfo {author}
  {\bibfnamefont {P.}~\bibnamefont {Stebler}}, \bibinfo {author} {\bibfnamefont
  {U.-J.}\ \bibnamefont {Wiese}},\ and\ \bibinfo {author} {\bibfnamefont
  {P.}~\bibnamefont {Zoller}},\ }\bibfield  {title} {\bibinfo {title} {Atomic
  quantum simulation of dynamical gauge fields coupled to fermionic matter:
  From string breaking to evolution after a quench},\ }\href
  {https://doi.org/10.1103/PhysRevLett.109.175302} {\bibfield  {journal}
  {\bibinfo  {journal} {Phys. Rev. Lett.}\ }\textbf {\bibinfo {volume} {109}},\
  \bibinfo {pages} {175302} (\bibinfo {year} {2012})},\ \Eprint
  {https://arxiv.org/abs/1205.6366} {arXiv:1205.6366} \BibitemShut {NoStop}%
\bibitem [{\citenamefont {Wiese}(2013)}]{Wiese2013}%
  \BibitemOpen
  \bibfield  {author} {\bibinfo {author} {\bibfnamefont {U.-J.}\ \bibnamefont
  {Wiese}},\ }\bibfield  {title} {\bibinfo {title} {Ultracold quantum gases and
  lattice systems: quantum simulation of lattice gauge theories},\ }\href
  {https://doi.org/10.1002/andp.201300104} {\bibfield  {journal} {\bibinfo
  {journal} {Ann. Phys. (L.)}\ }\textbf {\bibinfo {volume} {525}},\ \bibinfo
  {pages} {777} (\bibinfo {year} {2013})},\ \Eprint
  {https://arxiv.org/abs/1305.1602} {arXiv:1305.1602} \BibitemShut {NoStop}%
\bibitem [{\citenamefont {Zohar}\ \emph {et~al.}(2015)\citenamefont {Zohar},
  \citenamefont {Cirac},\ and\ \citenamefont {Reznik}}]{ZCR2015}%
  \BibitemOpen
  \bibfield  {author} {\bibinfo {author} {\bibfnamefont {E.}~\bibnamefont
  {Zohar}}, \bibinfo {author} {\bibfnamefont {J.~I.}\ \bibnamefont {Cirac}},\
  and\ \bibinfo {author} {\bibfnamefont {B.}~\bibnamefont {Reznik}},\
  }\bibfield  {title} {\bibinfo {title} {Quantum simulations of lattice gauge
  theories using ultracold atoms in optical lattices},\ }\href
  {https://doi.org/10.1088%2F0034-4885%2F79%2F1%2F014401} {\bibfield  {journal}
  {\bibinfo  {journal} {Rep. Prog. Phys.}\ }\textbf {\bibinfo {volume} {79}},\
  \bibinfo {pages} {014401} (\bibinfo {year} {2015})},\ \Eprint
  {https://arxiv.org/abs/1503.02312} {arXiv:1503.02312} \BibitemShut {NoStop}%
\bibitem [{\citenamefont {Weimer}\ \emph {et~al.}(2010)\citenamefont {Weimer},
  \citenamefont {M\"{u}ller}, \citenamefont {Lesanovsky}, \citenamefont
  {Zoller},\ and\ \citenamefont {B\"{u}chler}}]{WMLZplus2010}%
  \BibitemOpen
  \bibfield  {author} {\bibinfo {author} {\bibfnamefont {H.}~\bibnamefont
  {Weimer}}, \bibinfo {author} {\bibfnamefont {M.}~\bibnamefont {M\"{u}ller}},
  \bibinfo {author} {\bibfnamefont {I.}~\bibnamefont {Lesanovsky}}, \bibinfo
  {author} {\bibfnamefont {P.}~\bibnamefont {Zoller}},\ and\ \bibinfo {author}
  {\bibfnamefont {H.~P.}\ \bibnamefont {B\"{u}chler}},\ }\bibfield  {title}
  {\bibinfo {title} {A {R}ydberg quantum simulator},\ }\href
  {https://doi.org/10.1038/nphys1614} {\bibfield  {journal} {\bibinfo
  {journal} {Nature Phys.}\ }\textbf {\bibinfo {volume} {6}},\ \bibinfo {pages}
  {382} (\bibinfo {year} {2010})},\ \Eprint {https://arxiv.org/abs/0907.1657}
  {arXiv:0907.1657} \BibitemShut {NoStop}%
\bibitem [{\citenamefont {Tagliacozzo}\ \emph {et~al.}(2013)\citenamefont
  {Tagliacozzo}, \citenamefont {Celi}, \citenamefont {Zamora},\ and\
  \citenamefont {Lewenstein}}]{TCZL2013}%
  \BibitemOpen
  \bibfield  {author} {\bibinfo {author} {\bibfnamefont {L.}~\bibnamefont
  {Tagliacozzo}}, \bibinfo {author} {\bibfnamefont {A.}~\bibnamefont {Celi}},
  \bibinfo {author} {\bibfnamefont {A.}~\bibnamefont {Zamora}},\ and\ \bibinfo
  {author} {\bibfnamefont {M.}~\bibnamefont {Lewenstein}},\ }\bibfield  {title}
  {\bibinfo {title} {Optical {A}belian lattice gauge theories},\ }\href
  {https://doi.org/https://doi.org/10.1016/j.aop.2012.11.009} {\bibfield
  {journal} {\bibinfo  {journal} {Ann. Phys. (N. Y.)}\ }\textbf {\bibinfo
  {volume} {330}},\ \bibinfo {pages} {160} (\bibinfo {year} {2013})},\ \Eprint
  {https://arxiv.org/abs/1205.0496} {arXiv:1205.0496} \BibitemShut {NoStop}%
\bibitem [{\citenamefont {Byrnes}\ and\ \citenamefont
  {Yamamoto}(2006)}]{Byrnes2006}%
  \BibitemOpen
  \bibfield  {author} {\bibinfo {author} {\bibfnamefont {T.}~\bibnamefont
  {Byrnes}}\ and\ \bibinfo {author} {\bibfnamefont {Y.}~\bibnamefont
  {Yamamoto}},\ }\bibfield  {title} {\bibinfo {title} {Simulating lattice gauge
  theories on a quantum computer},\ }\href
  {https://doi.org/10.1103%2Fphysreva.73.022328} {\bibfield  {journal}
  {\bibinfo  {journal} {Phys. Rev. A}\ }\textbf {\bibinfo {volume} {73}},\
  \bibinfo {pages} {022328} (\bibinfo {year} {2006})},\ \Eprint
  {https://arxiv.org/abs/quant-ph/0510027} {arXiv:quant-ph/0510027}
  \BibitemShut {NoStop}%
\bibitem [{\citenamefont {Jordan}\ \emph {et~al.}(2012)\citenamefont {Jordan},
  \citenamefont {Lee},\ and\ \citenamefont {Preskill}}]{JLP2012}%
  \BibitemOpen
  \bibfield  {author} {\bibinfo {author} {\bibfnamefont {S.~P.}\ \bibnamefont
  {Jordan}}, \bibinfo {author} {\bibfnamefont {K.~S.~M.}\ \bibnamefont {Lee}},\
  and\ \bibinfo {author} {\bibfnamefont {J.}~\bibnamefont {Preskill}},\
  }\bibfield  {title} {\bibinfo {title} {Quantum algorithms for quantum field
  theories},\ }\href {https://doi.org/10.1126/science.1217069} {\bibfield
  {journal} {\bibinfo  {journal} {Science}\ }\textbf {\bibinfo {volume}
  {336}},\ \bibinfo {pages} {1130} (\bibinfo {year} {2012})},\ \Eprint
  {https://arxiv.org/abs/1111.3633} {arXiv:1111.3633} \BibitemShut {NoStop}%
\bibitem [{\citenamefont {Jordan}\ \emph
  {et~al.}(2014{\natexlab{a}})\citenamefont {Jordan}, \citenamefont {Lee},\
  and\ \citenamefont {Preskill}}]{JLP2014}%
  \BibitemOpen
  \bibfield  {author} {\bibinfo {author} {\bibfnamefont {S.~P.}\ \bibnamefont
  {Jordan}}, \bibinfo {author} {\bibfnamefont {K.~S.~M.}\ \bibnamefont {Lee}},\
  and\ \bibinfo {author} {\bibfnamefont {J.}~\bibnamefont {Preskill}},\
  }\bibfield  {title} {\bibinfo {title} {Quantum computation of scattering in
  scalar quantum field theories},\ }\href
  {https://dl.acm.org/doi/10.5555/2685155.2685163} {\bibfield  {journal}
  {\bibinfo  {journal} {Quantum Inf. Comput.}\ }\textbf {\bibinfo {volume}
  {14}},\ \bibinfo {pages} {1014–1080} (\bibinfo {year}
  {2014}{\natexlab{a}})},\ \Eprint {https://arxiv.org/abs/1112.4833}
  {arXiv:1112.4833} \BibitemShut {NoStop}%
\bibitem [{\citenamefont {Jordan}\ \emph
  {et~al.}(2014{\natexlab{b}})\citenamefont {Jordan}, \citenamefont {Lee},\
  and\ \citenamefont {Preskill}}]{JLP2014b}%
  \BibitemOpen
  \bibfield  {author} {\bibinfo {author} {\bibfnamefont {S.~P.}\ \bibnamefont
  {Jordan}}, \bibinfo {author} {\bibfnamefont {K.~S.~M.}\ \bibnamefont {Lee}},\
  and\ \bibinfo {author} {\bibfnamefont {J.}~\bibnamefont {Preskill}},\
  }\bibfield  {title} {\bibinfo {title} {Quantum algorithms for fermionic
  quantum field theories},\ }\href {https://arxiv.org/abs/1404.7115} {\bibfield
   {journal} {\bibinfo  {journal} {arXiv:1404.7115}\ } (\bibinfo {year}
  {2014}{\natexlab{b}})}\BibitemShut {NoStop}%
\bibitem [{\citenamefont {Brennen}\ \emph {et~al.}(2015)\citenamefont
  {Brennen}, \citenamefont {Rohde}, \citenamefont {Sanders},\ and\
  \citenamefont {Singh}}]{BRSS2015}%
  \BibitemOpen
  \bibfield  {author} {\bibinfo {author} {\bibfnamefont {G.~K.}\ \bibnamefont
  {Brennen}}, \bibinfo {author} {\bibfnamefont {P.}~\bibnamefont {Rohde}},
  \bibinfo {author} {\bibfnamefont {B.~C.}\ \bibnamefont {Sanders}},\ and\
  \bibinfo {author} {\bibfnamefont {S.}~\bibnamefont {Singh}},\ }\bibfield
  {title} {\bibinfo {title} {Multiscale quantum simulation of quantum field
  theory using wavelets},\ }\href {https://doi.org/10.1103/PhysRevA.92.032315}
  {\bibfield  {journal} {\bibinfo  {journal} {Phys. Rev. A}\ }\textbf {\bibinfo
  {volume} {92}},\ \bibinfo {pages} {032315} (\bibinfo {year} {2015})},\
  \Eprint {https://arxiv.org/abs/1412.0750} {arXiv:1412.0750} \BibitemShut
  {NoStop}%
\bibitem [{\citenamefont {Marshall}\ \emph {et~al.}(2015)\citenamefont
  {Marshall}, \citenamefont {Pooser}, \citenamefont {Siopsis},\ and\
  \citenamefont {Weedbrook}}]{MPSW2015}%
  \BibitemOpen
  \bibfield  {author} {\bibinfo {author} {\bibfnamefont {K.}~\bibnamefont
  {Marshall}}, \bibinfo {author} {\bibfnamefont {R.}~\bibnamefont {Pooser}},
  \bibinfo {author} {\bibfnamefont {G.}~\bibnamefont {Siopsis}},\ and\ \bibinfo
  {author} {\bibfnamefont {C.}~\bibnamefont {Weedbrook}},\ }\bibfield  {title}
  {\bibinfo {title} {Quantum simulation of quantum field theory using
  continuous variables},\ }\href {https://doi.org/10.1103/PhysRevA.92.063825}
  {\bibfield  {journal} {\bibinfo  {journal} {Phys. Rev. A}\ }\textbf {\bibinfo
  {volume} {92}},\ \bibinfo {pages} {063825} (\bibinfo {year} {2015})},\
  \Eprint {https://arxiv.org/abs/1503.08121} {arXiv:1503.08121} \BibitemShut
  {NoStop}%
\bibitem [{\citenamefont {Jordan}\ \emph {et~al.}(2018)\citenamefont {Jordan},
  \citenamefont {Krovi}, \citenamefont {Lee},\ and\ \citenamefont
  {Preskill}}]{JKLP2018}%
  \BibitemOpen
  \bibfield  {author} {\bibinfo {author} {\bibfnamefont {S.~P.}\ \bibnamefont
  {Jordan}}, \bibinfo {author} {\bibfnamefont {H.}~\bibnamefont {Krovi}},
  \bibinfo {author} {\bibfnamefont {K.~S.~M.}\ \bibnamefont {Lee}},\ and\
  \bibinfo {author} {\bibfnamefont {J.}~\bibnamefont {Preskill}},\ }\bibfield
  {title} {\bibinfo {title} {{BQP}-completeness of scattering in scalar quantum
  field theory},\ }\href {https://doi.org/10.22331/q-2018-01-08-44} {\bibfield
  {journal} {\bibinfo  {journal} {Quantum}\ }\textbf {\bibinfo {volume} {2}},\
  \bibinfo {pages} {44} (\bibinfo {year} {2018})},\ \Eprint
  {https://arxiv.org/abs/1703.00454} {arXiv:1703.00454} \BibitemShut {NoStop}%
\bibitem [{\citenamefont {Alexandru}\ \emph {et~al.}(2019)\citenamefont
  {Alexandru}, \citenamefont {Bedaque}, \citenamefont {Lamm},\ and\
  \citenamefont {Lawrence}}]{ABLL2019}%
  \BibitemOpen
  \bibfield  {author} {\bibinfo {author} {\bibfnamefont {A.}~\bibnamefont
  {Alexandru}}, \bibinfo {author} {\bibfnamefont {P.~F.}\ \bibnamefont
  {Bedaque}}, \bibinfo {author} {\bibfnamefont {H.}~\bibnamefont {Lamm}},\ and\
  \bibinfo {author} {\bibfnamefont {S.}~\bibnamefont {Lawrence}} (\bibinfo
  {collaboration} {NuQS Collaboration}),\ }\bibfield  {title} {\bibinfo {title}
  {$\ensuremath{\sigma}$ models on quantum computers},\ }\href
  {https://doi.org/10.1103/PhysRevLett.123.090501} {\bibfield  {journal}
  {\bibinfo  {journal} {Phys. Rev. Lett.}\ }\textbf {\bibinfo {volume} {123}},\
  \bibinfo {pages} {090501} (\bibinfo {year} {2019})},\ \Eprint
  {https://arxiv.org/abs/1903.06577} {arXiv:1903.06577} \BibitemShut {NoStop}%
\bibitem [{\citenamefont {Moosavian}\ \emph {et~al.}(2019)\citenamefont
  {Moosavian}, \citenamefont {Garrison},\ and\ \citenamefont
  {Jordan}}]{MGJ2019}%
  \BibitemOpen
  \bibfield  {author} {\bibinfo {author} {\bibfnamefont {A.~H.}\ \bibnamefont
  {Moosavian}}, \bibinfo {author} {\bibfnamefont {J.~R.}\ \bibnamefont
  {Garrison}},\ and\ \bibinfo {author} {\bibfnamefont {S.~P.}\ \bibnamefont
  {Jordan}},\ }\bibfield  {title} {\bibinfo {title} {Site-by-site quantum state
  preparation algorithm for preparing vacua of fermionic lattice field
  theories},\ }\href {https://arxiv.org/abs/1911.03505} {\bibfield  {journal}
  {\bibinfo  {journal} {arXiv:1911.03505}\ } (\bibinfo {year}
  {2019})}\BibitemShut {NoStop}%
\bibitem [{\citenamefont {Kanwar}\ and\ \citenamefont {Wagman}(2021)}]{KW21}%
  \BibitemOpen
  \bibfield  {author} {\bibinfo {author} {\bibfnamefont {G.}~\bibnamefont
  {Kanwar}}\ and\ \bibinfo {author} {\bibfnamefont {M.~L.}\ \bibnamefont
  {Wagman}},\ }\bibfield  {title} {\bibinfo {title} {Real-time lattice gauge
  theory actions: Unitarity, convergence, and path integral contour
  deformations},\ }\href {https://doi.org/10.1103/PhysRevD.104.014513}
  {\bibfield  {journal} {\bibinfo  {journal} {Phys. Rev. D}\ }\textbf {\bibinfo
  {volume} {104}},\ \bibinfo {pages} {014513} (\bibinfo {year} {2021})},\
  \Eprint {https://arxiv.org/abs/2103.02602} {arXiv:2103.02602} \BibitemShut
  {NoStop}%
\bibitem [{\citenamefont {Dalmonte}\ and\ \citenamefont
  {Montangero}(2016)}]{DM2016}%
  \BibitemOpen
  \bibfield  {author} {\bibinfo {author} {\bibfnamefont {M.}~\bibnamefont
  {Dalmonte}}\ and\ \bibinfo {author} {\bibfnamefont {S.}~\bibnamefont
  {Montangero}},\ }\bibfield  {title} {\bibinfo {title} {Lattice gauge theory
  simulations in the quantum information era},\ }\href
  {https://doi.org/10.1080/00107514.2016.1151199} {\bibfield  {journal}
  {\bibinfo  {journal} {Contemp. Phys.}\ }\textbf {\bibinfo {volume} {57}},\
  \bibinfo {pages} {388} (\bibinfo {year} {2016})},\ \Eprint
  {https://arxiv.org/abs/1602.03776} {arXiv:1602.03776} \BibitemShut {NoStop}%
\bibitem [{\citenamefont {{Ba{\~{n}}uls et al.}}(2020)}]{Banuls2020}%
  \BibitemOpen
  \bibfield  {author} {\bibinfo {author} {\bibfnamefont {M.~C.}\ \bibnamefont
  {{Ba{\~{n}}uls et al.}}},\ }\bibfield  {title} {\bibinfo {title} {Simulating
  lattice gauge theories within quantum technologies},\ }\href
  {https://doi.org/10.1140/epjd/e2020-100571-8} {\bibfield  {journal} {\bibinfo
   {journal} {Eur. Phys. J. D}\ }\textbf {\bibinfo {volume} {74}},\ \bibinfo
  {pages} {165} (\bibinfo {year} {2020})},\ \Eprint
  {https://arxiv.org/abs/1911.00003} {arXiv:1911.00003} \BibitemShut {NoStop}%
\bibitem [{\citenamefont {Ba{\~{n}}uls}\ and\ \citenamefont
  {Cichy}(2020)}]{BC2020}%
  \BibitemOpen
  \bibfield  {author} {\bibinfo {author} {\bibfnamefont {M.~C.}\ \bibnamefont
  {Ba{\~{n}}uls}}\ and\ \bibinfo {author} {\bibfnamefont {K.}~\bibnamefont
  {Cichy}},\ }\bibfield  {title} {\bibinfo {title} {Review on novel methods for
  lattice gauge theories},\ }\href {https://doi.org/10.1088/1361-6633/ab6311}
  {\bibfield  {journal} {\bibinfo  {journal} {Rep. Prog. Phys.}\ }\textbf
  {\bibinfo {volume} {83}},\ \bibinfo {pages} {024401} (\bibinfo {year}
  {2020})},\ \Eprint {https://arxiv.org/abs/1910.00257} {arXiv:1910.00257}
  \BibitemShut {NoStop}%
\bibitem [{\citenamefont {{Aidelsburger et al.}}(2021)}]{Aidelsburger2021}%
  \BibitemOpen
  \bibfield  {author} {\bibinfo {author} {\bibfnamefont {M.}~\bibnamefont
  {{Aidelsburger et al.}}},\ }\bibfield  {title} {\bibinfo {title} {Cold atoms
  meet lattice gauge theory},\ }\href
  {https://royalsocietypublishing.org/doi/10.1098/rsta.2021.0064} {\bibfield
  {journal} {\bibinfo  {journal} {Phil. Trans. R. Soc. A}\ }\textbf {\bibinfo
  {volume} {380}},\ \bibinfo {pages} {0064} (\bibinfo {year} {2021})},\ \Eprint
  {https://arxiv.org/abs/2106.03063} {arXiv:2106.03063} \BibitemShut {NoStop}%
\bibitem [{\citenamefont {Zohar}(2021)}]{Zohar2021}%
  \BibitemOpen
  \bibfield  {author} {\bibinfo {author} {\bibfnamefont {E.}~\bibnamefont
  {Zohar}},\ }\bibfield  {title} {\bibinfo {title} {Quantum simulation of
  lattice gauge theories in more than one space dimension—requirements,
  challenges and methods},\ }\href
  {https://royalsocietypublishing.org/doi/full/10.1098/rsta.2021.0069}
  {\bibfield  {journal} {\bibinfo  {journal} {Phil. Trans. R. Soc. A}\ }\textbf
  {\bibinfo {volume} {380}},\ \bibinfo {pages} {0069} (\bibinfo {year}
  {2021})},\ \Eprint {https://arxiv.org/abs/2106.04609} {arXiv:2106.04609}
  \BibitemShut {NoStop}%
\bibitem [{\citenamefont {Klco}\ \emph {et~al.}(2022)\citenamefont {Klco},
  \citenamefont {Roggero},\ and\ \citenamefont {Savage}}]{KRS2022}%
  \BibitemOpen
  \bibfield  {author} {\bibinfo {author} {\bibfnamefont {N.}~\bibnamefont
  {Klco}}, \bibinfo {author} {\bibfnamefont {A.}~\bibnamefont {Roggero}},\ and\
  \bibinfo {author} {\bibfnamefont {M.~J.}\ \bibnamefont {Savage}},\ }\bibfield
   {title} {\bibinfo {title} {Standard model physics and the digital quantum
  revolution: thoughts about the interface},\ }\href
  {https://doi.org/10.1088/1361-6633/ac58a4} {\bibfield  {journal} {\bibinfo
  {journal} {Rep. Prog. Phys.}\ }\textbf {\bibinfo {volume} {85}},\ \bibinfo
  {pages} {064301} (\bibinfo {year} {2022})},\ \Eprint
  {https://arxiv.org/abs/2107.04769} {arXiv:2107.04769} \BibitemShut {NoStop}%
\bibitem [{\citenamefont {Martinez}\ \emph {et~al.}(2016)\citenamefont
  {Martinez}, \citenamefont {Muschik}, \citenamefont {Schindler}, \citenamefont
  {Nigg}, \citenamefont {Erhard}, \citenamefont {Heyl}, \citenamefont {Hauke},
  \citenamefont {Dalmonte}, \citenamefont {Monz}, \citenamefont {Zoller},\ and\
  \citenamefont {Blatt}}]{Martinez16}%
  \BibitemOpen
  \bibfield  {author} {\bibinfo {author} {\bibfnamefont {E.~A.}\ \bibnamefont
  {Martinez}}, \bibinfo {author} {\bibfnamefont {C.~A.}\ \bibnamefont
  {Muschik}}, \bibinfo {author} {\bibfnamefont {P.}~\bibnamefont {Schindler}},
  \bibinfo {author} {\bibfnamefont {D.}~\bibnamefont {Nigg}}, \bibinfo {author}
  {\bibfnamefont {A.}~\bibnamefont {Erhard}}, \bibinfo {author} {\bibfnamefont
  {M.}~\bibnamefont {Heyl}}, \bibinfo {author} {\bibfnamefont {P.}~\bibnamefont
  {Hauke}}, \bibinfo {author} {\bibfnamefont {M.}~\bibnamefont {Dalmonte}},
  \bibinfo {author} {\bibfnamefont {T.}~\bibnamefont {Monz}}, \bibinfo {author}
  {\bibfnamefont {P.}~\bibnamefont {Zoller}},\ and\ \bibinfo {author}
  {\bibfnamefont {R.}~\bibnamefont {Blatt}},\ }\bibfield  {title} {\bibinfo
  {title} {Real-time dynamics of lattice gauge theories with a few-qubit
  quantum computer},\ }\href
  {https://www.nature.com/nature/journal/v534/n7608/full/nature18318.html}
  {\bibfield  {journal} {\bibinfo  {journal} {Nature}\ }\textbf {\bibinfo
  {volume} {534}},\ \bibinfo {pages} {516} (\bibinfo {year} {2016})},\ \Eprint
  {https://arxiv.org/abs/1605.04570} {arXiv:1605.04570} \BibitemShut {NoStop}%
\bibitem [{\citenamefont {Arrighi}(2019)}]{Arrighi2019}%
  \BibitemOpen
  \bibfield  {author} {\bibinfo {author} {\bibfnamefont {P.}~\bibnamefont
  {Arrighi}},\ }\bibfield  {title} {\bibinfo {title} {An overview of quantum
  cellular automata},\ }\href {https://doi.org/10.1007/s11047-019-09762-6}
  {\bibfield  {journal} {\bibinfo  {journal} {Natural Computing}\ }\textbf
  {\bibinfo {volume} {18}},\ \bibinfo {pages} {885} (\bibinfo {year} {2019})},\
  \Eprint {https://arxiv.org/abs/1904.12956} {arXiv:1904.12956} \BibitemShut
  {NoStop}%
\bibitem [{\citenamefont {Farrelly}(2020)}]{Farrelly2020}%
  \BibitemOpen
  \bibfield  {author} {\bibinfo {author} {\bibfnamefont {T.}~\bibnamefont
  {Farrelly}},\ }\bibfield  {title} {\bibinfo {title} {A review of quantum
  cellular automata},\ }\href {https://doi.org/10.22331/q-2020-11-30-368}
  {\bibfield  {journal} {\bibinfo  {journal} {Quantum}\ }\textbf {\bibinfo
  {volume} {4}},\ \bibinfo {pages} {368} (\bibinfo {year} {2020})},\ \Eprint
  {https://arxiv.org/abs/1904.13318} {arXiv:1904.13318} \BibitemShut {NoStop}%
\bibitem [{\citenamefont {Schumacher}\ and\ \citenamefont
  {Werner}(2004)}]{schumacher2004reversible}%
  \BibitemOpen
  \bibfield  {author} {\bibinfo {author} {\bibfnamefont {B.}~\bibnamefont
  {Schumacher}}\ and\ \bibinfo {author} {\bibfnamefont {R.~F.}\ \bibnamefont
  {Werner}},\ }\bibfield  {title} {\bibinfo {title} {Reversible quantum
  cellular automata},\ }\href@noop {} {\  (\bibinfo {year} {2004})},\ \Eprint
  {https://arxiv.org/abs/quant-ph/0405174} {arXiv:quant-ph/0405174}
  \BibitemShut {NoStop}%
\bibitem [{\citenamefont {Destri}\ and\ \citenamefont {{De
  Vega}}(1987)}]{DDV87}%
  \BibitemOpen
  \bibfield  {author} {\bibinfo {author} {\bibfnamefont {C.}~\bibnamefont
  {Destri}}\ and\ \bibinfo {author} {\bibfnamefont {H.~J.}\ \bibnamefont {{De
  Vega}}},\ }\bibfield  {title} {\bibinfo {title} {Light-cone lattice approach
  to fermionic theories in {2D}: The massive {T}hirring model},\ }\href
  {https://doi.org/https://doi.org/10.1016/0550-3213(87)90193-3} {\bibfield
  {journal} {\bibinfo  {journal} {Nucl. Phys. B}\ }\textbf {\bibinfo {volume}
  {290}},\ \bibinfo {pages} {363} (\bibinfo {year} {1987})}\BibitemShut
  {NoStop}%
\bibitem [{\citenamefont {Bisio}\ \emph {et~al.}(2018)\citenamefont {Bisio},
  \citenamefont {D'{A}riano}, \citenamefont {Perinotti},\ and\ \citenamefont
  {Tosini}}]{BDAPplus2018}%
  \BibitemOpen
  \bibfield  {author} {\bibinfo {author} {\bibfnamefont {A.}~\bibnamefont
  {Bisio}}, \bibinfo {author} {\bibfnamefont {G.~M.}\ \bibnamefont
  {D'{A}riano}}, \bibinfo {author} {\bibfnamefont {P.}~\bibnamefont
  {Perinotti}},\ and\ \bibinfo {author} {\bibfnamefont {A.}~\bibnamefont
  {Tosini}},\ }\bibfield  {title} {\bibinfo {title} {Thirring quantum cellular
  automaton},\ }\href {https://doi.org/10.1103/PhysRevA.97.032132} {\bibfield
  {journal} {\bibinfo  {journal} {Phys. Rev. A}\ }\textbf {\bibinfo {volume}
  {97}},\ \bibinfo {pages} {032132} (\bibinfo {year} {2018})},\ \Eprint
  {https://arxiv.org/abs/1711.03920} {arXiv:1711.03920} \BibitemShut {NoStop}%
\bibitem [{\citenamefont {Arrighi}\ \emph {et~al.}(2020)\citenamefont
  {Arrighi}, \citenamefont {B{\'{e}}ny},\ and\ \citenamefont
  {Farrelly}}]{ABF20}%
  \BibitemOpen
  \bibfield  {author} {\bibinfo {author} {\bibfnamefont {P.}~\bibnamefont
  {Arrighi}}, \bibinfo {author} {\bibfnamefont {C.}~\bibnamefont
  {B{\'{e}}ny}},\ and\ \bibinfo {author} {\bibfnamefont {T.}~\bibnamefont
  {Farrelly}},\ }\bibfield  {title} {\bibinfo {title} {A quantum cellular
  automaton for one-dimensional {QED}},\ }\href
  {https://doi.org/10.1007/s11128-019-2555-4} {\bibfield  {journal} {\bibinfo
  {journal} {Quantum Inf. Process.}\ }\textbf {\bibinfo {volume} {19}}
  (\bibinfo {year} {2020})},\ \Eprint {https://arxiv.org/abs/1903.07007}
  {arXiv:1903.07007} \BibitemShut {NoStop}%
\bibitem [{\citenamefont {Sellapillay}\ \emph {et~al.}(2022)\citenamefont
  {Sellapillay}, \citenamefont {Arrighi},\ and\ \citenamefont {{{Di}
  Molfetta}}}]{SADM2022}%
  \BibitemOpen
  \bibfield  {author} {\bibinfo {author} {\bibfnamefont {K.}~\bibnamefont
  {Sellapillay}}, \bibinfo {author} {\bibfnamefont {P.}~\bibnamefont
  {Arrighi}},\ and\ \bibinfo {author} {\bibfnamefont {G.}~\bibnamefont {{{Di}
  Molfetta}}},\ }\bibfield  {title} {\bibinfo {title} {A discrete relativistic
  spacetime formalism for 1+1-{QED} with continuum limits},\ }\href
  {https://www.nature.com/articles/s41598-022-06241-4#citeas} {\bibfield
  {journal} {\bibinfo  {journal} {Sci. Rep.}\ }\textbf {\bibinfo {volume}
  {12}},\ \bibinfo {pages} {2198} (\bibinfo {year} {2022})},\ \Eprint
  {https://arxiv.org/abs/2103.13150} {arXiv:2103.13150} \BibitemShut {NoStop}%
\bibitem [{\citenamefont {Eon}\ \emph {et~al.}(2022)\citenamefont {Eon},
  \citenamefont {{{D}i Molfetta}}, \citenamefont {Magnifico},\ and\
  \citenamefont {Arrighi}}]{EDMMplus22}%
  \BibitemOpen
  \bibfield  {author} {\bibinfo {author} {\bibfnamefont {N.}~\bibnamefont
  {Eon}}, \bibinfo {author} {\bibfnamefont {G.}~\bibnamefont {{{D}i
  Molfetta}}}, \bibinfo {author} {\bibfnamefont {G.}~\bibnamefont
  {Magnifico}},\ and\ \bibinfo {author} {\bibfnamefont {P.}~\bibnamefont
  {Arrighi}},\ }\bibfield  {title} {\bibinfo {title} {A relativistic discrete
  spacetime formulation of 3+1 {QED}},\ }\href@noop {} {\  (\bibinfo {year}
  {2022})},\ \Eprint {https://arxiv.org/abs/2205.03148} {arXiv:2205.03148}
  \BibitemShut {NoStop}%
\bibitem [{\citenamefont {Farrelly}(2015)}]{Farrelly15}%
  \BibitemOpen
  \bibfield  {author} {\bibinfo {author} {\bibfnamefont {T.~C.}\ \bibnamefont
  {Farrelly}},\ }\emph {\bibinfo {title} {Insights from quantum information
  into fundamental physics}},\ \href@noop {} {Ph.D. thesis},\ \bibinfo
  {school} {{U}niversity of {C}ambridge} (\bibinfo {year} {2015}),\ \Eprint
  {https://arxiv.org/abs/1708.08897} {arXiv:1708.08897} \BibitemShut {NoStop}%
\bibitem [{\citenamefont {Farrelly}\ and\ \citenamefont
  {Streich}(2020)}]{FS2020}%
  \BibitemOpen
  \bibfield  {author} {\bibinfo {author} {\bibfnamefont {T.}~\bibnamefont
  {Farrelly}}\ and\ \bibinfo {author} {\bibfnamefont {J.}~\bibnamefont
  {Streich}},\ }\bibfield  {title} {\bibinfo {title} {Discretizing quantum
  field theories for quantum simulation},\ }\href@noop {} {\  (\bibinfo {year}
  {2020})},\ \Eprint {https://arxiv.org/abs/2002.02643} {arXiv:2002.02643}
  \BibitemShut {NoStop}%
\bibitem [{\citenamefont {Yepez}(2016)}]{Yepez2016}%
  \BibitemOpen
  \bibfield  {author} {\bibinfo {author} {\bibfnamefont {J.}~\bibnamefont
  {Yepez}},\ }\bibfield  {title} {\bibinfo {title} {Quantum lattice gas
  algorithmic representation of gauge field theory},\ }in\ \href
  {https://doi.org/10.1117%2F12.2246702} {\emph {\bibinfo {booktitle} {Quantum
  Inf. Sci. Technol. {II}}}}\ (\bibinfo  {publisher} {{SPIE}-Intl Soc. Optical
  Eng.},\ \bibinfo {year} {2016})\ \Eprint {https://arxiv.org/abs/1609.02225}
  {arXiv:1609.02225} \BibitemShut {NoStop}%
\bibitem [{\citenamefont {Cedzich}\ \emph {et~al.}()\citenamefont {Cedzich},
  \citenamefont {Joye}, \citenamefont {Werner},\ and\ \citenamefont
  {Werner}}]{CJWW}%
  \BibitemOpen
  \bibfield  {author} {\bibinfo {author} {\bibfnamefont {C.}~\bibnamefont
  {Cedzich}}, \bibinfo {author} {\bibfnamefont {A.}~\bibnamefont {Joye}},
  \bibinfo {author} {\bibfnamefont {A.~H.}\ \bibnamefont {Werner}},\ and\
  \bibinfo {author} {\bibfnamefont {R.~F.}\ \bibnamefont {Werner}},\ }\bibfield
   {title} {\bibinfo {title} {Exponential tail estimates for quantum lattice
  dynamics},\ }\href@noop {} {\ }\bibinfo {note} {In preparation}\BibitemShut
  {NoStop}%
\bibitem [{\citenamefont {Vogts}(2013)}]{Vogts09}%
  \BibitemOpen
  \bibfield  {author} {\bibinfo {author} {\bibfnamefont {H.}~\bibnamefont
  {Vogts}},\ }\bibfield  {title} {\bibinfo {title} {\emph{Discrete Time Quantum
  Lattice Systems}},\ }\href
  {https://publikationsserver.tu-braunschweig.de/servlets/MCRFileNodeServlet/digibib_derivate_00007120/dissertation.pdf}
  {\bibfield  {journal} {\bibinfo  {journal} {Ph.D. thesis, Technischen
  Universit{\"a}t zu Braunschweig}\ } (\bibinfo {year} {2013})}\BibitemShut
  {NoStop}%
\bibitem [{\citenamefont {Bialynicki-Birula}(1994)}]{BB94a}%
  \BibitemOpen
  \bibfield  {author} {\bibinfo {author} {\bibfnamefont {I.}~\bibnamefont
  {Bialynicki-Birula}},\ }\bibfield  {title} {\bibinfo {title} {Weyl, {D}irac,
  and {M}axwell equations on a lattice as unitary cellular automata},\ }\href
  {https://journals.aps.org/prd/abstract/10.1103/PhysRevD.49.6920} {\bibfield
  {journal} {\bibinfo  {journal} {Phys. Rev. D}\ }\textbf {\bibinfo {volume}
  {49}},\ \bibinfo {pages} {6920} (\bibinfo {year} {1994})}\BibitemShut
  {NoStop}%
\bibitem [{\citenamefont {Farrelly}\ and\ \citenamefont
  {Short}(2014{\natexlab{a}})}]{Farrelly2014a}%
  \BibitemOpen
  \bibfield  {author} {\bibinfo {author} {\bibfnamefont {T.~C.}\ \bibnamefont
  {Farrelly}}\ and\ \bibinfo {author} {\bibfnamefont {A.~J.}\ \bibnamefont
  {Short}},\ }\bibfield  {title} {\bibinfo {title} {Discrete spacetime and
  relativistic quantum particles},\ }\href
  {https://doi.org/10.1103/physreva.89.062109} {\bibfield  {journal} {\bibinfo
  {journal} {Phys. Rev. A}\ }\textbf {\bibinfo {volume} {89}} (\bibinfo {year}
  {2014}{\natexlab{a}})},\ \Eprint {https://arxiv.org/abs/1312.2852}
  {arXiv:1312.2852} \BibitemShut {NoStop}%
\bibitem [{\citenamefont {Bisio}\ \emph {et~al.}(2015)\citenamefont {Bisio},
  \citenamefont {D'Ariano},\ and\ \citenamefont {Tosini}}]{Bisio2015}%
  \BibitemOpen
  \bibfield  {author} {\bibinfo {author} {\bibfnamefont {A.}~\bibnamefont
  {Bisio}}, \bibinfo {author} {\bibfnamefont {G.~M.}\ \bibnamefont
  {D'Ariano}},\ and\ \bibinfo {author} {\bibfnamefont {A.}~\bibnamefont
  {Tosini}},\ }\bibfield  {title} {\bibinfo {title} {Quantum field as a quantum
  cellular automaton: {T}he {D}irac free evolution in one dimension},\ }\href
  {https://doi.org/10.1016/j.aop.2014.12.016} {\bibfield  {journal} {\bibinfo
  {journal} {Ann. Phys. (N. Y.)}\ }\textbf {\bibinfo {volume} {354}},\ \bibinfo
  {pages} {244} (\bibinfo {year} {2015})},\ \Eprint
  {https://arxiv.org/abs/1212.2839} {arXiv:1212.2839} \BibitemShut {NoStop}%
\bibitem [{\citenamefont {Arnault}\ and\ \citenamefont
  {Debbasch}(2016)}]{AD16b}%
  \BibitemOpen
  \bibfield  {author} {\bibinfo {author} {\bibfnamefont {P.}~\bibnamefont
  {Arnault}}\ and\ \bibinfo {author} {\bibfnamefont {F.}~\bibnamefont
  {Debbasch}},\ }\bibfield  {title} {\bibinfo {title} {Quantum walks and
  discrete gauge theories},\ }\href
  {https://doi.org/10.1103/physreva.93.052301} {\bibfield  {journal} {\bibinfo
  {journal} {Phys. Rev. A}\ }\textbf {\bibinfo {volume} {93}},\ \bibinfo
  {pages} {052301} (\bibinfo {year} {2016})},\ \Eprint
  {https://arxiv.org/abs/1508.00038} {arXiv:1508.00038} \BibitemShut {NoStop}%
\bibitem [{\citenamefont {M{\'{a}}rquez-Mart{\'{\i}}n}\ \emph
  {et~al.}(2018)\citenamefont {M{\'{a}}rquez-Mart{\'{\i}}n}, \citenamefont
  {Arnault}, \citenamefont {{Di Molfetta}},\ and\ \citenamefont
  {P{\'{e}}rez}}]{MMAMP18}%
  \BibitemOpen
  \bibfield  {author} {\bibinfo {author} {\bibfnamefont {I.}~\bibnamefont
  {M{\'{a}}rquez-Mart{\'{\i}}n}}, \bibinfo {author} {\bibfnamefont
  {P.}~\bibnamefont {Arnault}}, \bibinfo {author} {\bibfnamefont
  {G.}~\bibnamefont {{Di Molfetta}}},\ and\ \bibinfo {author} {\bibfnamefont
  {A.}~\bibnamefont {P{\'{e}}rez}},\ }\bibfield  {title} {\bibinfo {title}
  {Electromagnetic lattice gauge invariance in two-dimensional discrete-time
  quantum walks},\ }\href {https://doi.org/10.1103/physreva.98.032333}
  {\bibfield  {journal} {\bibinfo  {journal} {Phys. Rev. A}\ }\textbf {\bibinfo
  {volume} {98}},\ \bibinfo {pages} {032333} (\bibinfo {year} {2018})},\
  \Eprint {https://arxiv.org/abs/1808.04488} {arXiv:1808.04488} \BibitemShut
  {NoStop}%
\bibitem [{\citenamefont {Cedzich}\ \emph {et~al.}(2019)\citenamefont
  {Cedzich}, \citenamefont {Geib}, \citenamefont {Werner},\ and\ \citenamefont
  {Werner}}]{CGWW18}%
  \BibitemOpen
  \bibfield  {author} {\bibinfo {author} {\bibfnamefont {C.}~\bibnamefont
  {Cedzich}}, \bibinfo {author} {\bibfnamefont {T.}~\bibnamefont {Geib}},
  \bibinfo {author} {\bibfnamefont {A.~H.}\ \bibnamefont {Werner}},\ and\
  \bibinfo {author} {\bibfnamefont {R.~F.}\ \bibnamefont {Werner}},\ }\bibfield
   {title} {\bibinfo {title} {Quantum walks in external gauge fields},\ }\href
  {https://doi.org/10.1063/1.5054894} {\bibfield  {journal} {\bibinfo
  {journal} {J. Math. Phys.}\ }\textbf {\bibinfo {volume} {60}},\ \bibinfo
  {pages} {012107} (\bibinfo {year} {2019})},\ \Eprint
  {https://arxiv.org/abs/1808.10850} {arXiv:1808.10850} \BibitemShut {NoStop}%
\bibitem [{\citenamefont {Arnault}\ \emph {et~al.}(2016)\citenamefont
  {Arnault}, \citenamefont {{Di Molfetta}}, \citenamefont {Brachet},\ and\
  \citenamefont {Debbasch}}]{AMBD16}%
  \BibitemOpen
  \bibfield  {author} {\bibinfo {author} {\bibfnamefont {P.}~\bibnamefont
  {Arnault}}, \bibinfo {author} {\bibfnamefont {G.}~\bibnamefont {{Di
  Molfetta}}}, \bibinfo {author} {\bibfnamefont {M.}~\bibnamefont {Brachet}},\
  and\ \bibinfo {author} {\bibfnamefont {F.}~\bibnamefont {Debbasch}},\
  }\bibfield  {title} {\bibinfo {title} {Quantum walks and non-{A}belian
  discrete gauge theory},\ }\href {https://doi.org/10.1103/physreva.94.012335}
  {\bibfield  {journal} {\bibinfo  {journal} {Phys. Rev. A}\ }\textbf {\bibinfo
  {volume} {94}},\ \bibinfo {pages} {012335} (\bibinfo {year} {2016})},\
  \Eprint {https://arxiv.org/abs/1605.01605} {arXiv:1605.01605} \BibitemShut
  {NoStop}%
\bibitem [{\citenamefont {{Di Molfetta}}\ \emph {et~al.}(2013)\citenamefont
  {{Di Molfetta}}, \citenamefont {Brachet},\ and\ \citenamefont
  {Debbasch}}]{DMD13b}%
  \BibitemOpen
  \bibfield  {author} {\bibinfo {author} {\bibfnamefont {G.}~\bibnamefont {{Di
  Molfetta}}}, \bibinfo {author} {\bibfnamefont {M.}~\bibnamefont {Brachet}},\
  and\ \bibinfo {author} {\bibfnamefont {F.}~\bibnamefont {Debbasch}},\
  }\bibfield  {title} {\bibinfo {title} {Quantum walks as massless {D}irac
  fermions in curved space},\ }\href
  {https://link.aps.org/doi/10.1103/PhysRevA.88.042301} {\bibfield  {journal}
  {\bibinfo  {journal} {Phys. Rev. A}\ }\textbf {\bibinfo {volume} {88}},\
  \bibinfo {pages} {042301} (\bibinfo {year} {2013})},\ \Eprint
  {https://arxiv.org/abs/1212.5821} {arXiv:1212.5821} \BibitemShut {NoStop}%
\bibitem [{\citenamefont {{Di Molfetta}}\ \emph {et~al.}(2014)\citenamefont
  {{Di Molfetta}}, \citenamefont {Debbasch},\ and\ \citenamefont
  {Brachet}}]{DMD14}%
  \BibitemOpen
  \bibfield  {author} {\bibinfo {author} {\bibfnamefont {G.}~\bibnamefont {{Di
  Molfetta}}}, \bibinfo {author} {\bibfnamefont {F.}~\bibnamefont {Debbasch}},\
  and\ \bibinfo {author} {\bibfnamefont {M.}~\bibnamefont {Brachet}},\
  }\bibfield  {title} {\bibinfo {title} {Quantum walks in artificial electric
  and gravitational fields},\ }\href
  {http://www.sciencedirect.com/science/article/pii/S0378437113011059}
  {\bibfield  {journal} {\bibinfo  {journal} {Physica A}\ }\textbf {\bibinfo
  {volume} {397}},\ \bibinfo {pages} {157} (\bibinfo {year} {2014})},\ \Eprint
  {https://arxiv.org/abs/1309.4923} {arXiv:1309.4923} \BibitemShut {NoStop}%
\bibitem [{\citenamefont {Arnault}\ and\ \citenamefont
  {Debbasch}(2017)}]{AD17}%
  \BibitemOpen
  \bibfield  {author} {\bibinfo {author} {\bibfnamefont {P.}~\bibnamefont
  {Arnault}}\ and\ \bibinfo {author} {\bibfnamefont {F.}~\bibnamefont
  {Debbasch}},\ }\bibfield  {title} {\bibinfo {title} {Quantum walks and
  gravitational waves},\ }\href {https://doi.org/10.1016/j.aop.2017.04.003}
  {\bibfield  {journal} {\bibinfo  {journal} {Ann. Phys.}\ }\textbf {\bibinfo
  {volume} {383}},\ \bibinfo {pages} {645} (\bibinfo {year} {2017})},\ \Eprint
  {https://arxiv.org/abs/1609.00722} {arXiv:1609.00722} \BibitemShut {NoStop}%
\bibitem [{\citenamefont {Arrighi}\ \emph {et~al.}(2016)\citenamefont
  {Arrighi}, \citenamefont {Facchini},\ and\ \citenamefont
  {Forets}}]{Arrighi_curved_1D_15}%
  \BibitemOpen
  \bibfield  {author} {\bibinfo {author} {\bibfnamefont {P.}~\bibnamefont
  {Arrighi}}, \bibinfo {author} {\bibfnamefont {S.}~\bibnamefont {Facchini}},\
  and\ \bibinfo {author} {\bibfnamefont {M.}~\bibnamefont {Forets}},\
  }\bibfield  {title} {\bibinfo {title} {Quantum walking in curved spacetime},\
  }\href {https://link.springer.com/article/10.1007/s11128-016-1335-7}
  {\bibfield  {journal} {\bibinfo  {journal} {Quantum Inf. Process.}\ }\textbf
  {\bibinfo {volume} {15}},\ \bibinfo {pages} {3467} (\bibinfo {year}
  {2016})},\ \Eprint {https://arxiv.org/abs/1505.07023} {arXiv:1505.07023}
  \BibitemShut {NoStop}%
\bibitem [{\citenamefont {Arrighi}\ and\ \citenamefont
  {Facchini}(2017)}]{AF17}%
  \BibitemOpen
  \bibfield  {author} {\bibinfo {author} {\bibfnamefont {P.}~\bibnamefont
  {Arrighi}}\ and\ \bibinfo {author} {\bibfnamefont {S.}~\bibnamefont
  {Facchini}},\ }\bibfield  {title} {\bibinfo {title} {Quantum walking in
  curved spacetime: (3+1) dimensions, and beyond},\ }\href
  {http://dl.acm.org/citation.cfm?id=3179561.3179565} {\bibfield  {journal}
  {\bibinfo  {journal} {Quantum Inf. Comput.}\ }\textbf {\bibinfo {volume}
  {17}},\ \bibinfo {pages} {810} (\bibinfo {year} {2017})},\ \Eprint
  {https://arxiv.org/abs/1609.00305} {arXiv:1609.00305} \BibitemShut {NoStop}%
\bibitem [{\citenamefont {Arnault}(2017)}]{Arnault2017}%
  \BibitemOpen
  \bibfield  {author} {\bibinfo {author} {\bibfnamefont {P.}~\bibnamefont
  {Arnault}},\ }\emph {\bibinfo {title} {Discrete-time quantum walk and gauge
  theories}},\ \href@noop {} {Ph.D. thesis},\ \bibinfo  {school}
  {Universit{\'e} Pierre et Marie Curie} (\bibinfo {year} {2017}),\ \Eprint
  {https://arxiv.org/abs/1710.11123} {arXiv:1710.11123} \BibitemShut {NoStop}%
\bibitem [{\citenamefont {Arrighi}\ \emph {et~al.}(2019)\citenamefont
  {Arrighi}, \citenamefont {{Di Molfetta}}, \citenamefont {Marquez-Martin},\
  and\ \citenamefont {Perez}}]{ADMMMplus2019}%
  \BibitemOpen
  \bibfield  {author} {\bibinfo {author} {\bibfnamefont {P.}~\bibnamefont
  {Arrighi}}, \bibinfo {author} {\bibfnamefont {G.}~\bibnamefont {{Di
  Molfetta}}}, \bibinfo {author} {\bibfnamefont {I.}~\bibnamefont
  {Marquez-Martin}},\ and\ \bibinfo {author} {\bibfnamefont {A.}~\bibnamefont
  {Perez}},\ }\bibfield  {title} {\bibinfo {title} {From curved spacetime to
  spacetime-dependent local unitaries over the honeycomb and triangular quantum
  walks},\ }\href {https://doi.org/10.1038/s41598-019-47535-4} {\bibfield
  {journal} {\bibinfo  {journal} {Sci. Rep.}\ }\textbf {\bibinfo {volume}
  {9}},\ \bibinfo {pages} {10904} (\bibinfo {year} {2019})},\ \Eprint
  {https://arxiv.org/abs/1812.02601} {arXiv:1812.02601} \BibitemShut {NoStop}%
\bibitem [{\citenamefont {Farrelly}\ and\ \citenamefont
  {Short}(2014{\natexlab{b}})}]{Farrelly2014b}%
  \BibitemOpen
  \bibfield  {author} {\bibinfo {author} {\bibfnamefont {T.~C.}\ \bibnamefont
  {Farrelly}}\ and\ \bibinfo {author} {\bibfnamefont {A.~J.}\ \bibnamefont
  {Short}},\ }\bibfield  {title} {\bibinfo {title} {Causal fermions in discrete
  space-time},\ }\href {https://doi.org/10.1103/physreva.89.012302} {\bibfield
  {journal} {\bibinfo  {journal} {Phys. Rev. A}\ }\textbf {\bibinfo {volume}
  {89}} (\bibinfo {year} {2014}{\natexlab{b}})},\ \Eprint
  {https://arxiv.org/abs/1303.4652} {arXiv:1303.4652} \BibitemShut {NoStop}%
\bibitem [{\citenamefont {D'Ariano}\ and\ \citenamefont
  {Perinotti}(2016)}]{DAP16}%
  \BibitemOpen
  \bibfield  {author} {\bibinfo {author} {\bibfnamefont {G.~M.}\ \bibnamefont
  {D'Ariano}}\ and\ \bibinfo {author} {\bibfnamefont {P.}~\bibnamefont
  {Perinotti}},\ }\bibfield  {title} {\bibinfo {title} {Quantum cellular
  automata and free quantum field theory},\ }\href
  {https://doi.org/10.1007/s11467-016-0616-z} {\bibfield  {journal} {\bibinfo
  {journal} {Front. Phys.}\ }\textbf {\bibinfo {volume} {12}},\ \bibinfo
  {pages} {120301} (\bibinfo {year} {2016})},\ \Eprint
  {https://arxiv.org/abs/1608.02004} {arXiv:1608.02004} \BibitemShut {NoStop}%
\bibitem [{\citenamefont {Debbasch}(2019{\natexlab{a}})}]{Debbasch2019a}%
  \BibitemOpen
  \bibfield  {author} {\bibinfo {author} {\bibfnamefont {F.}~\bibnamefont
  {Debbasch}},\ }\bibfield  {title} {\bibinfo {title} {Action principles for
  quantum automata and {L}orentz invariance of discrete time quantum walks},\
  }\href {https://doi.org/10.1016/j.aop.2019.03.005} {\bibfield  {journal}
  {\bibinfo  {journal} {Ann. Phys.}\ }\textbf {\bibinfo {volume} {405}},\
  \bibinfo {pages} {340} (\bibinfo {year} {2019}{\natexlab{a}})},\ \Eprint
  {https://arxiv.org/abs/1806.02313} {arXiv:1806.02313} \BibitemShut {NoStop}%
\bibitem [{\citenamefont {Arnault}(2022)}]{Arnault2022}%
  \BibitemOpen
  \bibfield  {author} {\bibinfo {author} {\bibfnamefont {P.}~\bibnamefont
  {Arnault}},\ }\bibfield  {title} {\bibinfo {title} {Clifford algebra from
  quantum automata and unitary {W}ilson fermions},\ }\href
  {https://doi.org/10.1103/PhysRevA.106.012201} {\bibfield  {journal} {\bibinfo
   {journal} {Phys. Rev. A}\ }\textbf {\bibinfo {volume} {106}},\ \bibinfo
  {pages} {012201} (\bibinfo {year} {2022})},\ \Eprint
  {https://arxiv.org/abs/2105.12314} {arXiv:2105.12314} \BibitemShut {NoStop}%
\bibitem [{\citenamefont {Susskind}(1977)}]{Susskind77a}%
  \BibitemOpen
  \bibfield  {author} {\bibinfo {author} {\bibfnamefont {L.}~\bibnamefont
  {Susskind}},\ }\bibfield  {title} {\bibinfo {title} {Lattice fermions},\
  }\href {https://doi.org/10.1103/physrevd.16.3031} {\bibfield  {journal}
  {\bibinfo  {journal} {Phys. Rev. D}\ }\textbf {\bibinfo {volume} {16}},\
  \bibinfo {pages} {3031} (\bibinfo {year} {1977})}\BibitemShut {NoStop}%
\bibitem [{\citenamefont {Arnault}\ \emph {et~al.}(2020)\citenamefont
  {Arnault}, \citenamefont {Pepper},\ and\ \citenamefont
  {P{\'{e}}rez}}]{APP20}%
  \BibitemOpen
  \bibfield  {author} {\bibinfo {author} {\bibfnamefont {P.}~\bibnamefont
  {Arnault}}, \bibinfo {author} {\bibfnamefont {B.}~\bibnamefont {Pepper}},\
  and\ \bibinfo {author} {\bibfnamefont {A.}~\bibnamefont {P{\'{e}}rez}},\
  }\bibfield  {title} {\bibinfo {title} {Quantum walks in weak electric fields
  and {B}loch oscillations},\ }\href
  {https://doi.org/10.1103/physreva.101.062324} {\bibfield  {journal} {\bibinfo
   {journal} {Phys. Rev. A}\ }\textbf {\bibinfo {volume} {101}},\ \bibinfo
  {pages} {062324} (\bibinfo {year} {2020})},\ \Eprint
  {https://arxiv.org/abs/2001.05346} {arXiv:2001.05346} \BibitemShut {NoStop}%
\bibitem [{\citenamefont {Bachmann}(2007)}]{SB05_notes}%
  \BibitemOpen
  \bibfield  {author} {\bibinfo {author} {\bibfnamefont {S.}~\bibnamefont
  {Bachmann}},\ }\href
  {https://www.epfl.ch/schools/sb/research/iphys/wp-content/uploads/2018/10/QFT_main.pdf}
  {\emph {\bibinfo {title} {Champs Quantiques Relativistes}}}\ (\bibinfo {year}
  {2007})\ \bibinfo {note} {notes du cours du Professeur Mikha{\"i}l
  Shaposhnikov}\BibitemShut {NoStop}%
\bibitem [{\citenamefont {Cedzich}\ \emph {et~al.}(2013)\citenamefont
  {Cedzich}, \citenamefont {Ryb{\'a}r}, \citenamefont {Werner}, \citenamefont
  {Alberti}, \citenamefont {Genske},\ and\ \citenamefont {Werner}}]{ced13}%
  \BibitemOpen
  \bibfield  {author} {\bibinfo {author} {\bibfnamefont {C.}~\bibnamefont
  {Cedzich}}, \bibinfo {author} {\bibfnamefont {T.}~\bibnamefont {Ryb{\'a}r}},
  \bibinfo {author} {\bibfnamefont {A.~H.}\ \bibnamefont {Werner}}, \bibinfo
  {author} {\bibfnamefont {A.}~\bibnamefont {Alberti}}, \bibinfo {author}
  {\bibfnamefont {M.}~\bibnamefont {Genske}},\ and\ \bibinfo {author}
  {\bibfnamefont {R.~F.}\ \bibnamefont {Werner}},\ }\bibfield  {title}
  {\bibinfo {title} {Propagation of quantum walks in electric fields},\ }\href
  {https://link.aps.org/doi/10.1103/PhysRevLett.111.160601} {\bibfield
  {journal} {\bibinfo  {journal} {Phys. Rev. Lett.}\ }\textbf {\bibinfo
  {volume} {111}},\ \bibinfo {pages} {160601} (\bibinfo {year} {2013})},\
  \Eprint {https://arxiv.org/abs/1302.2081} {arXiv:1302.2081} \BibitemShut
  {NoStop}%
\bibitem [{\citenamefont {Cedzich}\ and\ \citenamefont {Werner}(2021)}]{CW19}%
  \BibitemOpen
  \bibfield  {author} {\bibinfo {author} {\bibfnamefont {C.}~\bibnamefont
  {Cedzich}}\ and\ \bibinfo {author} {\bibfnamefont {A.~H.}\ \bibnamefont
  {Werner}},\ }\bibfield  {title} {\bibinfo {title} {Anderson localization for
  electric quantum walks and skew-shift {CMV} matrices},\ }\href
  {https://link.springer.com/article/10.1007/s00220-021-04204-w} {\bibfield
  {journal} {\bibinfo  {journal} {Comm. Math. Phys.}\ }\textbf {\bibinfo
  {volume} {387}},\ \bibinfo {pages} {1257} (\bibinfo {year} {2021})},\ \Eprint
  {https://arxiv.org/abs/1906.11931} {arXiv:1906.11931} \BibitemShut {NoStop}%
\bibitem [{\citenamefont {Christiansen}\ and\ \citenamefont
  {Halvorsen}(2009)}]{CH09}%
  \BibitemOpen
  \bibfield  {author} {\bibinfo {author} {\bibfnamefont {S.~H.}\ \bibnamefont
  {Christiansen}}\ and\ \bibinfo {author} {\bibfnamefont {T.~G.}\ \bibnamefont
  {Halvorsen}},\ }\bibfield  {title} {\bibinfo {title} {Convergence of lattice
  gauge theory for {M}axwell's equations},\ }\href
  {https://doi.org/10.1007/s10543-009-0242-z} {\bibfield  {journal} {\bibinfo
  {journal} {{BIT} Numerical Mathematics}\ }\textbf {\bibinfo {volume} {49}},\
  \bibinfo {pages} {645} (\bibinfo {year} {2009})}\BibitemShut {NoStop}%
\bibitem [{\citenamefont {Arrighi}\ \emph {et~al.}(2014)\citenamefont
  {Arrighi}, \citenamefont {Facchini},\ and\ \citenamefont {Forets}}]{AFF14a}%
  \BibitemOpen
  \bibfield  {author} {\bibinfo {author} {\bibfnamefont {P.}~\bibnamefont
  {Arrighi}}, \bibinfo {author} {\bibfnamefont {S.}~\bibnamefont {Facchini}},\
  and\ \bibinfo {author} {\bibfnamefont {M.}~\bibnamefont {Forets}},\
  }\bibfield  {title} {\bibinfo {title} {Discrete {L}orentz covariance for
  quantum walks and quantum cellular automata},\ }\href
  {https://iopscience.iop.org/article/10.1088/1367-2630/16/9/093007} {\bibfield
   {journal} {\bibinfo  {journal} {New. {J}. {P}hys.}\ }\textbf {\bibinfo
  {volume} {16}},\ \bibinfo {pages} {093007} (\bibinfo {year} {2014})},\
  \Eprint {https://arxiv.org/abs/1404.4499} {arXiv:1404.4499} \BibitemShut
  {NoStop}%
\bibitem [{\citenamefont {Debbasch}(2019{\natexlab{b}})}]{Debbasch2019b}%
  \BibitemOpen
  \bibfield  {author} {\bibinfo {author} {\bibfnamefont {F.}~\bibnamefont
  {Debbasch}},\ }\bibfield  {title} {\bibinfo {title} {Discrete geometry from
  quantum walks},\ }\href {https://doi.org/10.3390/condmat4020040} {\bibfield
  {journal} {\bibinfo  {journal} {Condensed Matter}\ }\textbf {\bibinfo
  {volume} {4}},\ \bibinfo {pages} {40} (\bibinfo {year}
  {2019}{\natexlab{b}})},\ \Eprint {https://arxiv.org/abs/1902.11079}
  {arXiv:1902.11079} \BibitemShut {NoStop}%
\end{thebibliography}

%

\clearpage

\newpage

\appendix

\section{Variational principle for unitary  fermions}
\label{app:variational_principle}

For brevity, in this appendix we write the Hamiltonian of Eq.\ \eqref{eq:HDQW} as
\begin{equation}
\mathcal{H}^{\text{DQW}} = \mathcal{H} \, .
\end{equation}
For the sake of pedagogy, we are going to consider here the variational problem for the equations of motion, with the non-real action, i.e., that of Eq.\ \eqref{eq:action}, because with it one gets a good feeling of what is going on, but below in Appendix \ref{app:Euler-Lagrange} we will also consider this variational problem on the generic real action of Eq.\ \eqref{eq:Lagrangian_density} (a particular case of which is the real action of Eq.\ \eqref{eq:action_symmetric}).

The action of Eq.\ \eqref{eq:action} can be rewritten in the following Hamiltonian form,
\begin{subequations}
\begin{align}
S_{\text{DQW}}^{\text{asym.}} &= \epsilon \sum_j \langle \psi_j \, | \, \mathrm{i} d_{0} \psi \big|_j - \mathcal{H} \psi_j \rangle \label{eq:action_2}\\ 
&\equiv \epsilon^2 \sum_j \sum_p \psi^{\dag}_{j,p} \left[ \mathrm{i} d_0 \psi \big|_{j,p} - (\mathcal{H} \psi_j)_p \right] \, ,
\end{align}
\end{subequations}
where $d_0$ is the symmetric lattice derivative defined in Eq.\ \eqref{eq:discretize}, and where we have introduced the Hermitian product
\begin{equation}
\langle \psi_j \, | \, \phi_j \rangle  \, \defeq \epsilon \sum_{p}  \psi^{\dag}_{j,p} \phi_{j,p} \, .
\end{equation}
We remark that here and below $\mathcal H$ could depend on time, but we omit the corresponding index $j$ to keep the notation light.

The action Eq.\ \eqref{eq:action_2} is a functional of the following real independent fields, $\text{Re}(\psi^a)$ and $\text{Im}(\psi^a)$, where $a=1,2$ labels the internal components of $\psi$.
In the continuum, one can show that variational problems for a Dirac-field action can be equivalently solved by considering this action as a functional of the following complex independent fields, $\psi^a$ and $(\psi^{a})^{\ast}$, $a=1,2$. We assume that this feature is robust to the discretization. Moreover, since the discretization is the same for all internal components of the field, for convenience we treat the fields collectively (see Sec.\ \ref{subsec:framework}) as $\psi$ and  $\psi^{\dag}$, and therefore write
\begin{equation}
S_{\text{DQW}}^{\text{asym.}} \equiv \mathscr{S}[\psi,\psi^{\dag}] \, .
\end{equation}

Let $\psi$ be a field that extremalizes $S_{\text{DQW}}^{\text{asym.}}$, and consider an arbitrary variation $\delta \psi$ of $\psi$.
Let us compute the variation of the action under the variations $\delta\psi$ and $\delta\psi^{\dag}$, namely,
\begin{equation}
\delta S \defeq \mathscr{S}[\psi+\delta\psi, \psi^{\dag} + \delta \psi^{\dag}] - \mathscr{S}[\psi, \psi^{\dag}]  \, .
\end{equation}
In variational problems like this, one can usually show that as long as the action is linear in its arguments, one can keep in $\delta S$ only the first-order terms in the variations $\delta\psi$ and $\delta\psi^{\dag}$, which we are going to assume and which here delivers,
\begin{align}
\delta S &= \epsilon^2 \sum_{j,p} \delta \psi^{\dag}_{j,p} \left[ \frac{
\mathrm{i}}{2\epsilon} (\psi_{j+1,p} - \psi_{j-1,p}) - (\mathcal{H} \psi_j)_p \right] \nonumber \\
& \quad+  \epsilon^2 \sum_{j,p} \psi^{\dag}_{j,p} \left[\frac{
\mathrm{i}}{2\epsilon} (\delta \psi_{j+1,p} - \delta \psi_{j-1,p}) - (\mathcal{H} \delta \psi_j)_p \right] \\
&=: \delta S_1 + \delta S_2 \, ,
\end{align}
where
\begin{subequations}
\begin{align}
\delta S_1 &\defeq \epsilon \sum_j \Big\langle  \delta \psi_j \, \Big| \, \frac{\mathrm{i}}{2\epsilon} ( \psi_{j+1,p} -  \psi_{j-1,p}) - \mathcal{H}\psi_j \Big\rangle  \\
\delta S_2 &\defeq \epsilon\sum_j \Big\langle  \psi_j \, \Big| \, \frac{\mathrm{i}}{2\epsilon} ( \delta \psi_{j+1,p} -  \delta \psi_{j-1,p}) - \mathcal{H} \delta \psi_j \Big\rangle   \, .\label{eq:deltaS2}
\end{align}
\end{subequations}

If we choose $\delta \psi = 0$, then $\delta S = \delta S_1$, and requiring $\delta S = 0$ for any $\delta \psi^{\dag}$ trivially delivers the correct equations of motion (EOM).
To see that this also works if $\delta \psi^{\dag} = 0$, so that $\delta S = \delta S_2$, we need to do some work on $\delta S_2$.
First of all, if $j$ runs from an initial $j_i$ to a final $j_f$ and $p$ from $p_i$ to $p_f$, then because of the symmetric lattice derivatives the action $S_{\text{DQW}}^{\text{asym.}}$ in Eq.\ \eqref{eq:action_2} must be defined with a $\sum_{j=j_i+1}^{j_f-1}$ and a $\sum_{p=p_i+1}^{p_f-1}$, i.e., omitting the first and last indices.
Then, from \eqref{eq:deltaS2} we have that
\begin{widetext}
\begin{align}
\delta S_2 &= \epsilon \sum_{j=j_i+1}^{j_f-1} \Big\langle  \psi_j \, \Big| \, \frac{\mathrm{i}}{2\epsilon} \delta \psi_{j+1} \Big\rangle  
+ \epsilon \sum_{j=j_i+1}^{j_f-1} \Big\langle  \psi_j \, \Big| \, - \frac{\mathrm{i}}{2\epsilon}  \delta \psi_{j-1} \Big\rangle 
+ \epsilon \sum_{j=j_i+1}^{j_f-1}  \Big\langle  \psi_j \, \Big| \, - \mathcal{H} \delta \psi_j \Big\rangle  \, .
\end{align}
After (i) performing a discrete integration by parts (i.e., a shift) in $j$ in the two first terms, and (ii) using the definition of the adjoint of $\mathcal{H}$ in the third one, we obtain
\begin{align}
\label{eq:sum}
\delta S_2 &= \epsilon  \sum_{j=j_i+2}^{j_f} \Big\langle - \frac{\mathrm{i}}{2\epsilon} \psi_{j-1} \, \Big| \, \delta \psi_{j} \Big\rangle 
+ \epsilon  \sum_{j=j_i}^{j_f-2} \Big\langle   \frac{\mathrm{i}}{2\epsilon}  \psi_{j+1} \, \Big| \,  \delta \psi_{j} \Big\rangle 
+ \epsilon  \sum_{j=j_i+1}^{j_f-1}  \Big\langle  - \mathcal{H}^{\dag} \psi_j \, \Big| \, \delta \psi_j \Big\rangle  \, .
\end{align}
At this point of the derivation of such a variational problem, it is frequent to say that we keep the ends of $\psi_j$ fixed, i.e., choose the lattice Neumann boundary conditions
\begin{equation}
\label{eq:fixed_ends}
\delta \psi_{j_i} = \delta \psi_{j_i+1} = \delta \psi_{j_f-1} = \delta \psi_{j_f} = 0 \, ,
\end{equation}
 so as not to have to deal with them. Note that this is not necessary for now, but it will be in the final reasoning where it ensures that the solution $\psi$ is determined, i.e., unique. 
With these boundary conditions the first term of the right-hand side of Eq.\ \eqref{eq:sum} becomes
\begin{subequations}
\begin{align}
\epsilon  \sum_{j=j_i+2}^{j_f} \Big\langle - \frac{\mathrm{i}}{2\epsilon} \psi_{j-1} \, \Big| \, \delta \psi_{j} \Big\rangle   \ &= \epsilon  \sum_{j=j_i+2}^{j_f-1} \Big\langle - \frac{\mathrm{i}}{2\epsilon} \psi_{j-1} \, \Big| \, \delta \psi_{j} \Big\rangle  + \Big\langle - \frac{\mathrm{i}}{2\epsilon} \psi_{N-1} \, \Big| \, \underbrace{\delta \psi_{N}}_{=0} \Big\rangle  + \Big\langle - \frac{\mathrm{i}}{2\epsilon} \psi_{0} \, \Big| \, \underbrace{  \delta \psi_{1}}_{=0}  \Big\rangle  \\
&= \epsilon  \sum_{j=j_i+1}^{j_f-1} \Big\langle - \frac{\mathrm{i}}{2\epsilon} \psi_{j-1} \, \Big| \, \delta \psi_{j} \Big\rangle  \, . \label{eq:term1}
\end{align}
\end{subequations}
Similarly, the the second term of the right-hand side of Eq.\ \eqref{eq:sum} becomes
\begin{align}
\epsilon  \sum_{j=j_i}^{j_f-2} \Big\langle  \frac{\mathrm{i}}{2\epsilon} \psi_{j+1} \, \Big| \, \delta \psi_{j} \Big\rangle 
&= \epsilon  \sum_{j=j_i+1}^{j_f-1} \Big\langle \frac{\mathrm{i}}{2\epsilon} \psi_{j+1} \, \Big| \, \delta \psi_{j} \Big\rangle  \, . \label{eq:term2}
\end{align}
\end{widetext}
Inserting Eqs.\ \eqref{eq:term1} and \eqref{eq:term2} into Eq.\ \eqref{eq:sum} and using the hermiticity of $\mathcal{H}$, we finally obtain that
\begin{equation}\label{eq:dS2}
\delta S_2 = \epsilon  \sum_{j=j_i+1}^{j_f-1} \Big\langle  \frac{\mathrm{i}}{2\epsilon} ( \psi_{j+1,p} -  \psi_{j-1,p}) - \mathcal{H}\psi_j \, \Big| \,   \delta \psi_j \Big\rangle  \, .
\end{equation}
If we choose $\delta \psi^{\dag} = 0$, then $\delta S = \delta S_2$, and requiring $\delta S = 0$ for any $\delta \psi$ now trivially delivers the correct EOM of Eq.\ \eqref{eq:DQW_fermions}. This completes the proof that the action of Eq.\ \eqref{eq:action} is a valid one for unitary fermions up to not being real-valued, an issue which is immediately solved by considering the real-valued action of Eq.\ \eqref{eq:action_symmetric} instead.

If we had not fixed the ends of $\psi_j$ we would have obtained \eqref{eq:dS2} with additional boundary terms. Then, to arrive at the EOM in the ``bulk'' one has to successively choose particular variation functions $\delta \psi_j$ as usual for this kind of proofs, with in the present case the (additional) requirement that all these variations functions have vanishing ends. Fixing the ends in the beginning like we did is thus not necessary but merely convenient, and we could have and would have obtained the same EOM if we had done the derivation with loose ends.

\section{Equivalence of the asymmetric and symmetric actions of unitary fermions}
\label{app:symmetric_action}

Let us show that the asymmetric action of unitary fermions, Eq.\ \eqref{eq:action}, is equal to the symmetric action of Eq.\ \eqref{eq:action_symmetric} up to boundary terms.
We start from the asymmetric action:
\begin{subequations}
\begin{align}
S_{\text{DQW}}^{\text{asym.}}  &= \epsilon^2 \sum_n \psi^{\dag}_n (\tilde{\gamma}^0)^{-1} \left[ \left( \mathrm{i} \tilde{\gamma}^{\mu} \frac{1}{2\epsilon} (\mathcal{T}^{-1}_{\mu} - \mathcal{T}_{\mu}) - {m} \right) \psi \right]_n \\
&=\epsilon \frac{\mathrm{i}}{2} \sum_n  \psi^{\dag}_n (\tilde{\gamma}^0)^{-1}  \tilde{\gamma}^{\mu} \psi_{n+\hat{\mu}} + \epsilon (\frac{-\mathrm{i}}{2}) \sum_n \psi^{\dag}_n (\tilde{\gamma}^0)^{-1} \tilde{\gamma}^{\mu} \psi_{n-\hat{\mu}} \nonumber \\
& \ \ \  - \epsilon \sum_n \epsilon {m} \psi^{\dag}_n (\tilde{\gamma}^0)^{-1} \psi_n  \label{eq:initial}
\end{align}
\end{subequations}
We consider the second term, and perform discrete integrations by parts on this term, i.e., a shift of one lattice site in $\mu$ for $\mu=0,1$.
To avoid cumbersome notations, we write explicitly $n=(j,p)$. Then, the second term of $S_{\text{DQW}}^{\text{asym.}}$ becomes
\begin{widetext}
\begin{subequations}
\begin{align}
T^{(2)} &\defeq \epsilon (\frac{-\mathrm{i}}{2}) \sum_{j=j_i+1}^{j_f-1} \sum_{p=p_i+1}^{p_f-1}  \psi^{\dag}_{j,p} (\tilde{\gamma}^0)^{-1} \tilde{\gamma}^{\mu}  \psi_{(j,p)-\hat{\mu}} \label{eq:T2}\\
&=\epsilon (\frac{-\mathrm{i}}{2}) \Bigg[ \sum_{j=j_i+1}^{j_f-1} \sum_{p=p_i+1}^{p_f-1}  \psi^{\dag}_{j,p} (\tilde{\gamma}^0)^{-1} \tilde{\gamma}^{0}  \psi_{j-1,p} + \sum_{j=j_i+1}^{j_f-1} \sum_{p=p_i+1}^{p_f-1}  \psi^{\dag}_{j,p} (\tilde{\gamma}^0)^{-1} \tilde{\gamma}^{1}  \psi_{j,p-1} \Bigg] \\
&= \epsilon (\frac{-\mathrm{i}}{2}) \Bigg[ \underbrace{\sum_{j=j_i}^{j_f-2} \sum_{p=p_i+1}^{p_f-1}  \psi^{\dag}_{j+1,p} (\tilde{\gamma}^0)^{-1} \tilde{\gamma}^{0}  \psi_{j,p}}_{A^{(0)}} + \underbrace{\sum_{j=j_i+1}^{j_f-1} \sum_{p=p_i}^{p_f-2}  \psi^{\dag}_{j,p+1} (\tilde{\gamma}^0)^{-1} \tilde{\gamma}^{1}  \psi_{j,p}}_{A^{(1)}} \Bigg]\, .
\label{eq:T2_3}
\end{align}
\end{subequations}
Extracting the boundary term $j=j_i$ out of $A^{(0)}$ and adding and effective zero yields
{\small
\begin{subequations}
\begin{align}
A^{(0)} &= \sum_{j=j_i+1}^{j_f-2} \sum_{p=p_i+1}^{p_f-1}  \psi^{\dag}_{j+1,p} (\tilde{\gamma}^0)^{-1} \tilde{\gamma}^{0}  \psi_{j,p} + \sum_{p=p_i+1}^{p_f-1}  \psi^{\dag}_{j_i+1,p} (\tilde{\gamma}^0)^{-1} \tilde{\gamma}^{0} \psi_{j_i,p} +  \underbrace{\sum_{p=p_i+1}^{p_f-1}  \psi^{\dag}_{j_f,p} (\tilde{\gamma}^0)^{-1} \tilde{\gamma}^{0} \psi_{j_f-1,p} - \sum_{p=p_i+1}^{p_f-1}  \psi^{\dag}_{j_f,p} (\tilde{\gamma}^0)^{-1} \tilde{\gamma}^{0} \psi_{j_f-1,p}}_{= 0} \\
&= \sum_{j=j_i+1}^{j_f-1} \sum_{p=p_i+1}^{p_f-1}  \psi^{\dag}_{j+1,p} (\tilde{\gamma}^0)^{-1} \tilde{\gamma}^{0} \psi_{j,p}  + \underbrace{ \sum_{p=p_i+1}^{p_f-1} \left( \psi^{\dag}_{j_i+1,p} (\tilde{\gamma}^0)^{-1} \tilde{\gamma}^{0} \psi_{j_i,p} - \psi^{\dag}_{j_f,p} (\tilde{\gamma}^0)^{-1} \tilde{\gamma}^{0} \psi_{j_f-1,p} \right)}_{B^{(0)}} \, . \label{eq:B0}
\end{align}
\end{subequations}}
Similarly, we obtain
\begin{align}
A^{(1)} &= \sum_{j=j_i+1}^{j_f-1} \sum_{p=p_i+1}^{p_f-1}  \psi^{\dag}_{j,p+1} (\tilde{\gamma}^0)^{-1} \tilde{\gamma}^{1}  \psi_{j,p}  +  \underbrace{\sum_{j=j_i+1}^{j_f-1} \left( \psi^{\dag}_{j,p_i+1} (\tilde{\gamma}^0)^{-1} \tilde{\gamma}^{1}  \psi_{j,p_i} - \psi^{\dag}_{j,p_f} (\tilde{\gamma}^0)^{-1} \tilde{\gamma}^{1}  \psi_{j,p_f-1} \right)}_{B^{(1)}} \, . \label{eq:B1}
\end{align}
\end{widetext}
Inserting these expressions, Eqs.\ \eqref{eq:B0} and \eqref{eq:B1}, back into \eqref{eq:T2_3}, we obtain
\begin{equation}
\label{eq:ekk}
T^{(2)} = \epsilon (\frac{-\mathrm{i}}{2})  \sum_n  \psi^{\dag}_{n+\hat{\mu}} (\tilde{\gamma}^0)^{-1} \tilde{\gamma}^{\mu}  \psi_{j,p} + B^{(0)} + B^{(1)} \, .
\end{equation}
Now, equating
\begin{align}
 (\tilde{\gamma}^0)^{-1} \tilde{\gamma}^{\mu} = \frac{\gamma^0}{\mu_{\epsilon}} \tilde{\gamma}^{\mu} \gamma^0 \gamma^0 
 = \gamma^0 \tilde{\gamma}^{\mu} \gamma^0 \frac{\gamma^0}{\mu_{\epsilon}} 
 = (\tilde{\gamma}^{\mu})^{\dag} ((\tilde{\gamma}^0)^{-1})^{\dag} \, ,
\end{align}
and inserting the last equality into Eq.\ \eqref{eq:ekk} delivers
\begin{align}
T^{(2)} = 
\epsilon (\frac{-\mathrm{i}}{2})  \sum_n  \psi^{\dag}_{n+\hat{\mu}} (\tilde{\gamma}^{\mu})^{\dag} ((\tilde{\gamma}^0)^{-1})^{\dag} \psi_{j,p} + B^{(0)} + B^{(1)} \, . \label{eq:finalT2}
\end{align}
Inserting this back into Eq.\ \eqref{eq:initial}, we obtain
\begin{equation}
S_{\text{DQW}}^{\text{asym.}}  = S_{\text{DQW}} + B^{(0)} + B^{(1)}  \, ,
\end{equation}
where $S_{\text{DQW}}$ has been defined by Eq.\ \eqref{eq:action_symmetric}.

The boundary term $B^{(0)} + B^{(1)}$ is irrelevant in the variational problem for determining the equations of motion.
Notice that it is possible to write the boundary term $B^{(0)} + B^{(1)}$ as a (lattice) $(1+1)$-divergence as it is usually done in the continuum, although for the current defined in that way to be real we have to consider the Hermitian conjugate part of the action: this is done in a more general framework in Sec.\ \ref{sec:Noether} and in Appendix \ref{app:Noether}. 

\section{Euler-Lagrange equations (from a real action)}
\label{app:Euler-Lagrange}

Consider the real action of Eq.\ \eqref{eq:Lagrangian_density}, and a field $\psi$ that extremalizes it.
Varying the field arbitrarily by $\delta \psi$ results in the action
\begin{widetext}
\begin{subequations}
\begin{align}
S_F' &\defeq \sum_n \mathscr{L}\left(\psi_n + \delta\psi_n, \psi_{n+\hat{\mu}} + \delta\psi_{n+\hat{\mu}}, \psi^{\dag}_n + \delta\psi^{\dag}_n, \psi^{\dag}_{n+\hat{\mu}}  + \delta\psi^{\dag}_{n+\hat{\mu}} \right)\\
&= S_F + \sum_n \left( \left. \frac{\partial\mathscr{L}}{\partial \psi_n} \right|_n \delta\psi_n + \left. \frac{\partial\mathscr{L}}{\partial \psi_{n+\hat{\mu}}} \right|_n \delta\psi_{n+\hat{\mu}} + \delta\psi^{\dag}_n \left. \frac{\partial\mathscr{L}}{\partial \psi^{\dag}_n} \right|_n +  \delta\psi^{\dag}_{n+\hat{\mu}} \left. \frac{\partial\mathscr{L}}{\partial \psi^{\dag}_{n+\hat{\mu}}} \right|_n  \right) \, , \label{eq:Sprime}
\end{align}
\end{subequations}
\end{widetext}
where in the second equation we have only kept terms up to first order since the action is linear in its arguments, and with implicit sums over $\mu$ in the second and fourth terms inside the brackets.
Eq.\ \eqref{eq:Sprime} can be rewritten as
\begin{equation}
\delta S \defeq S_F' - S_F = \delta S_1 + \delta S_2 \, ,
\end{equation}
where $\delta S_1$ is the variation of the fields and $\delta S_2$ that of the conjugate fields, i.e.
\begin{subequations}
\begin{align}
\delta S_1 &\defeq \sum_n  \left. \frac{\partial\mathscr{L}}{\partial \psi_n} \right|_n \delta\psi_n + \delta S_{11} \label{eq:deltaS1bis}\\
\delta S_2 &\defeq \sum_n  \left( \delta\psi^{\dag}_n \left. \frac{\partial\mathscr{L}}{\partial \psi^{\dag}_n} \right|_n +  \delta\psi^{\dag}_{n+\hat{\mu}} \left. \frac{\partial\mathscr{L}}{\partial \psi^{\dag}_{n+\hat{\mu}}} \right|_n  \right) \, ,
\end{align}
\end{subequations}
where
\begin{equation}
 \delta S_{11} \defeq \sum_n   \left. \frac{\partial\mathscr{L}}{\partial \psi_{n+\hat{\mu}}} \right|_n \delta\psi_{n+\hat{\mu}} \, .
\end{equation}
``Splitting time and space'' gives
\begin{equation}
\label{eq:deltaS11}
\delta S_{11} = \delta S_{11}^{(0)} + S_{11}^{(1)} \, ,
\end{equation}
where
\begin{subequations}
\begin{align}
\delta S_{11}^{(0)} &\defeq \sum_{j=j_i}^{j_f-1} \sum_{p=p_i}^{p_f-1} \left. \frac{\partial\mathscr{L}}{\partial \psi_{n+\hat{0}}} \right|_j \delta\psi_{j+1} \\
\delta S_{11}^{(1)} &\defeq \sum_{j=j_i}^{j_f-1} \sum_{p=p_i}^{p_f-1} \left. \frac{\partial\mathscr{L}}{\partial \psi_{n+\hat{1}}} \right|_p \delta\psi_{p+1} \, ,
\end{align}
\end{subequations}
where we have written down the index only when it is the ``working'' index, i.e. the index on which we are going to perform modifications.
Using the boundary conditions of Eq.\ \eqref{eq:fixed_ends}, we have that
\begin{widetext}
\begin{subequations}
\begin{align}
\delta S_{11}^{(0)} &=  \sum_{j=j_i+1}^{j_f} \sum_{p=p_i}^{p_f-1} \left. \frac{\partial\mathscr{L}}{\partial \psi_{n+\hat{0}}} \right|_{j-1} \delta\psi_{j}  \\
&= \sum_{j=j_i+1}^{j_f-1} \sum_{p=p_i}^{p_f-1} \left. \frac{\partial\mathscr{L}}{\partial \psi_{n+\hat{0}}} \right|_{j-1} \delta\psi_{j} +  \sum_{p=p_i}^{p_f-1} \left. \frac{\partial\mathscr{L}}{\partial \psi_{n+\hat{0}}} \right|_{j_f-1} \underbrace{\delta\psi_{j_f}}_{=0} + \sum_{p=p_i}^{p_f-1} \left. \frac{\partial\mathscr{L}}{\partial \psi_{n+\hat{0}}} \right|_{j_i-1} \underbrace{\delta\psi_{j_i}}_{=0}  - \sum_{p=p_i}^{p_f-1} \left. \frac{\partial\mathscr{L}}{\partial \psi_{n+\hat{0}}} \right|_{j_i-1} \underbrace{\delta\psi_{j_i}}_{=0} \\
&= \sum_{j=j_i}^{j_f-1} \sum_{p=p_i}^{p_f-1} \left. \frac{\partial\mathscr{L}}{\partial \psi_{n+\hat{0}}} \right|_{j-1} \delta\psi_{j} \, . \label{eq:firsteq}
\end{align}
\end{subequations}
\end{widetext}
Similarly, we can show that
\begin{equation}
\delta S_{11}^{(1)} = \sum_{j=j_i}^{j_f-1} \sum_{p=p_i}^{p_f-1} \left. \frac{\partial\mathscr{L}}{\partial \psi_{n+\hat{1}}} \right|_{p-1} \delta\psi_{p} \, , \label{eq:secondeq}
\end{equation}
so that in total we obtain
\begin{equation}
\delta S_{11} = \sum_n   \left. \frac{\partial\mathscr{L}}{\partial \psi_{n+\hat{\mu}}} \right|_{n-\hat{\mu}} \delta\psi_{n} \, ,
\end{equation}
with an implicit sum over $\mu$.
Inserting this into Eq.\ \eqref{eq:deltaS1bis} yields
\begin{equation}
    \delta S_1 =  \sum_n  \left( \left. \frac{\partial\mathscr{L}}{\partial \psi_n} \right|_n +  \left. \frac{\partial\mathscr{L}}{\partial \psi_{n+\hat{\mu}}} \right|_{n-\hat{\mu}} \right) \delta\psi_n \, .
\end{equation}
Similarly, we can show that
\begin{equation}
    \delta S_2 =  \sum_n   \delta\psi^{\dag}_n \left( \left. \frac{\partial\mathscr{L}}{\partial \psi^{\dag}_n} \right|_n +  \left. \frac{\partial\mathscr{L}}{\partial \psi^{\dag}_{n+\hat{\mu}}} \right|_{n-\hat{\mu}} \right) \, ,
\end{equation}
and so
\begin{align}
    \delta S &=  \sum_n  \left( \left. \frac{\partial\mathscr{L}}{\partial \psi_n} \right|_n +  \left. \frac{\partial\mathscr{L}}{\partial \psi_{n+\hat{\mu}}} \right|_{n-\hat{\mu}} \right) \delta\psi_n \nonumber \\
    &\quad+ \sum_n   \delta\psi^{\dag}_n \left( \left. \frac{\partial\mathscr{L}}{\partial \psi^{\dag}_n} \right|_n +  \left. \frac{\partial\mathscr{L}}{\partial \psi^{\dag}_{n+\hat{\mu}}} \right|_{n-\hat{\mu}} \right) \, .
\end{align}
Recall that $\psi$ extremalizes the action, i.e., $\delta S = 0$.
By choosing $\delta \psi_n = 0$, we must have $\delta S_2 = 0$ for any variation $\delta \psi^{\dag}$. 
Similarly, by choosing  $\delta \psi^{\dag}_n = 0$, implies that $\delta S_1 = 0$ for any variation $\delta \psi^{\dag}$.
Together, these conditions give the Euler-Lagrange equations
\begin{subequations}
\begin{align}
    \left. \frac{\partial\mathscr{L}}{\partial \psi_n} \right|_n +  \left. \frac{\partial\mathscr{L}}{\partial \psi_{n+\hat{\mu}}} \right|_{n-\hat{\mu}}&=0 \\
    \left. \frac{\partial\mathscr{L}}{\partial \psi^{\dag}_n} \right|_n +  \left. \frac{\partial\mathscr{L}}{\partial \psi^{\dag}_{n+\hat{\mu}}} \right|_{n-\hat{\mu}}&=0.
\end{align}
\end{subequations}
It might be possible to prove that for a real action these equations imply each other (more precisely, by proving that they are the Hermitian conjugates of each other), we leave it as an open problem.

\section{Proof of the lattice Noether's theorem for internal symmetries}
\label{app:Noether}

To the title of this appendix, we should actually add ``that depend only on a single real parameter $\alpha$'', but as mentioned in Sec.\ \ref{sec:Noether} the generalization to an arbitrary number of real parameters exhibits no major difficulty.

Under the transformation $\psi_n\longrightarrow \psi'_n = \psi_n + C_n \delta \alpha$ of Eq.\ \eqref{eq:field_transfo}, the action defined in Eq.\ \eqref{eq:Lagrangian_density} becomes
\begin{widetext}
\begin{subequations}
\begin{align}
S_F' &\defeq \sum_n \mathscr{L}(\psi'_n,\psi'_{n+\hat{\mu}}, {(\psi_n')}^{\dag}, (\psi_{n+\hat{\mu}}')^{\dag}) \\
&= \sum_n \mathscr{L}(\psi_n + C_n \delta \alpha, \ \psi_{n+\hat{\mu}} + C_{n+\hat{\mu}} \delta \alpha, \ \psi_n^{\dag} + C_n^{\dag} \delta \alpha , \ \psi_{n+\hat{\mu}}^{\dag} + C_{n+\hat{\mu}}^{\dag} \delta \alpha ) \\
&= \sum_n \mathscr{L}(\psi_n,\psi_{n+\hat{\mu}}, \psi_n^{\dag}, \psi_{n+\hat{\mu}}^{\dag}) + \left( \left. \frac{\partial\mathscr{L}}{\partial\psi_n}\right|_n C_n + \left. \frac{\partial\mathscr{L}}{\partial\psi_{n+\hat{\mu}}}\right|_n C_{n+\hat{\mu}} + C^{\dag}_n \left. \frac{\partial\mathscr{L}}{\partial\psi_n^{\dag}}\right|_n + C^{\dag}_{n+\hat{\mu}} \left. \frac{\partial\mathscr{L}}{\partial\psi_{n+\hat{\mu}}^{\dag}}\right|_n \right) \delta \alpha + O(\delta\alpha^2) \, ,
\end{align}
\end{subequations}
\end{widetext}
where we implicitly sum over $\mu$ in the second and fourth terms of the big bracket.
Hence, we obtain
\begin{equation}
\delta S \defeq S_F' - S_F = (Q_1 + Q_2)\delta \alpha \, ,
\label{eq:Q1plusQ2}
\end{equation}
where
{\small
\begin{subequations}
\begin{align}
Q_1 &\defeq \sum_n \left( \left. \frac{\partial\mathscr{L}}{\partial\psi_n}\right|_n C_n + C_n^{\dag} \left. \frac{\partial\mathscr{L}}{\partial\psi^{\dag}_n}\right|_n  \right) \\
Q_2 &\defeq \sum_n \left( \left. \frac{\partial\mathscr{L}}{\partial\psi_{n+\hat{\mu}}}\right|_n C_{n+\hat{\mu}} + C_{n+\hat{\mu}}^{\dag} \left. \frac{\partial\mathscr{L}}{\partial\psi^{\dag}_{n+\hat{\mu}}}\right|_n  \right) \, .
\end{align}
\end{subequations}}
In going from the first to the second equality in Eqs.\ \eqref{eq:Q1plusQ2}, we only kept, as in previous appendices, the first order in $\delta \alpha$, because the action is again assumed to be linear in its arguments.

We are going to perform a discrete integration by parts on $Q_2$.
To this end, let us ``split time and space'':
\begin{equation}
\label{eq:the_sum}
Q_2 = Q_2^{(0)} + Q_2^{(1)} \, ,
\end{equation}
with temporal and the spatial terms
{\small
\begin{subequations}
\begin{align}
Q_2^{(0)} &\defeq \sum_{j=j_i+1}^{j_f-1} \sum_{p=p_i+1}^{p_f-1} \left( \left. \frac{\partial\mathscr{L}}{\partial\psi_{n+\hat{0}}}\right|_j C_{j+1} + C_{j+1}^{\dag} \left. \frac{\partial\mathscr{L}}{\partial\psi^{\dag}_{n+\hat{0}}}\right|_j \right) \\
Q_2^{(1)} &\defeq  \sum_{j=j_i+1}^{j_f-1} \sum_{p=p_i+1}^{p_f-1} \left( \left. \frac{\partial\mathscr{L}}{\partial\psi_{n+\hat{1}}}\right|_p C_{p+1} + C_{p+1}^{\dag} \left. \frac{\partial\mathscr{L}}{\partial\psi^{\dag}_{n+\hat{1}}}\right|_p \right)  \, .
\end{align}
\end{subequations}}
Unless otherwise mentioned, for brevity and clarity we only write explicitly the coordinate at which we consider an expression whenever any of its factors is shifted, since this is the only coordinate on which ``we are going to work''.
Let us now perform a temporal discrete integration by parts on $Q_2^{(0)} $:
\begin{widetext}
{
\begin{subequations}
\begin{align}
Q_2^{(0)} &= \sum_{j=j_i+2}^{j_f} \sum_{p=p_i+1}^{p_f-1} \left( \left. \frac{\partial\mathscr{L}}{\partial\psi_{n+\hat{0}}}\right|_{j-1} C_{j} + C_{j}^{\dag} \left. \frac{\partial\mathscr{L}}{\partial\psi^{\dag}_{n+\hat{0}}}\right|_{j-1} \right) \\
&= \sum_{j=j_i+1}^{j_f-1} \sum_{p=p_i+1}^{p_f-1} \left( \left. \frac{\partial\mathscr{L}}{\partial\psi_{n+\hat{0}}}\right|_{j-1} C_{j} + C_{j}^{\dag} \left. \frac{\partial\mathscr{L}}{\partial\psi^{\dag}_{n+\hat{0}}}\right|_{j-1} \right) + b^{(0)} \, , \label{eq:first_term}
\end{align}
\end{subequations}}
where we have introduced the following boundary term,
\begin{equation}
b^{(0)} \defeq \sum_{p=p_i+1}^{p_f-1} \left[\left( \left. \frac{\partial\mathscr{L}}{\partial\psi_{n+\hat{0}}}\right|_{j_f-1} C_{j_f} + C_{j_f}^{\dag} \left. \frac{\partial\mathscr{L}}{\partial\psi^{\dag}_{n+\hat{0}}}\right|_{j_f-1} \right) - \left( \left. \frac{\partial\mathscr{L}}{\partial\psi_{n+\hat{0}}}\right|_{j_i} C_{j_i+1} + C_{j_i+1}^{\dag} \left. \frac{\partial\mathscr{L}}{\partial\psi^{\dag}_{n+\hat{0}}}\right|_{j_i} \right)\right] \, .
\end{equation}
This boundary term can be rewritten as the telescoping sum
\begin{equation}
b^{(0)}    = \sum_{p=p_i+1}^{p_f-1} \sum_{j=j_i+1}^{j_f-1}  \left( J^0_{j} - J^0_{j-1} \right) = \sum_n \epsilon \left(  d_0^L J^0\right)_{\! n} \, ,
\label{eq:boundary0}
\end{equation}
where $J^0$ has been defined in Eq.\ \eqref{eq:Kmu}.
Similarly, we perform a spatial discrete integration by parts on $Q_2^{(1)}$, which gives
\begin{equation}
Q_2^{(1)} = \sum_{j=j_i+1}^{j_f-1} \sum_{p=p_i+1}^{p_f-1} \left( \left. \frac{\partial\mathscr{L}}{\partial\psi_{n+\hat{1}}}\right|_{p-1} C_{p} + C_{p}^{\dag} \left. \frac{\partial\mathscr{L}}{\partial\psi^{\dag}_{n+\hat{1}}}\right|_{p-1} \right) + b^{(1)} \, , \label{eq:second_term}
\end{equation}
with boundary term
\begin{equation}
b^{(1)} \defeq \sum_{j=j_i+1}^{j_f-1} \left[\left( \left. \frac{\partial\mathscr{L}}{\partial\psi_{n+\hat{1}}}\right|_{p_f-1} C_{p_f} + C_{p_f}^{\dag} \left. \frac{\partial\mathscr{L}}{\partial\psi^{\dag}_{n+\hat{1}}}\right|_{p_f-1} \right) - \left( \left. \frac{\partial\mathscr{L}}{\partial\psi_{n+\hat{1}}}\right|_{p_i} C_{p_i+1} + C_{p_i+1}^{\dag} \left. \frac{\partial\mathscr{L}}{\partial\psi^{\dag}_{n+\hat{1}}}\right|_{p_i} \right)\right] \, .
\end{equation}
Again, this boundary term can be rewritten in terms of $J^1$ defined in Eq.\ \eqref{eq:Kmu} as
\begin{equation}
b^{(1)} =    \sum_{j=j_i+1}^{j_f-1} \sum_{p=p_i+1}^{p_f-1} \left( J^1_{p} - J^1_{p-1} \right)= \sum_n \epsilon \left( d_1^L J^1\right)_{\! n} \, .
\label{eq:boundary1}
\end{equation}

Inserting Eqs.\ \eqref{eq:first_term} and \eqref{eq:second_term} into Eq.\ \eqref{eq:the_sum}, and then the resulting expression into Eq.\ \eqref{eq:Q1plusQ2}, we obtain, after rearranging the first two sums and using the final expressions of the boundary terms, Eqs.\ \eqref{eq:boundary0} and \eqref{eq:boundary1}, the following expression,
\begin{equation}
\delta S =  \Bigg( \sum_n \Bigg[ \left( \left. \frac{\partial\mathscr{L}}{\partial\psi_n}\right|_n + \left. \frac{\partial\mathscr{L}}{\partial\psi_{n+\hat{\mu}}}\right|_{n-\hat{\mu}} \right) C_n + C_n^{\dag} \left( \left. \frac{\partial\mathscr{L}}{\partial\psi^{\dag}_n}\right|_n + \left. \frac{\partial\mathscr{L}}{\partial\psi^{\dag}_{n+\hat{\mu}}}\right|_{n-\hat{\mu}} \right)  \Bigg] + \sum_n \epsilon \left( d_{\mu}^L J^{\mu}\right)_{\! n} \Bigg) \delta \alpha \, .
\end{equation}
As in Appendix \ref{app:Euler-Lagrange} one can check that the first two summands are Euler-Lagrange expressions which vanish on shell\footnote{We leave it as an open problem whether for a real action these two Euler-Lagrange expressions are Hermitian conjugates of each other.}.
Thus, finally, we obtain that, on shell,
\begin{equation}
\label{eq:finall}
\delta S = \sum_n \epsilon \left( d_{\mu}^L J^{\mu} \right)_{\! n} \delta \alpha \, .
\end{equation}
If the considered transformation is a symmetry, i.e., if $\delta S = 0$, then this implies that $J$ is conserved on the lattice, i.e., Eq.\ \eqref{eq:1plus1divergence}, which ends the proof of our lattice Noether's theorem.

\end{widetext}

\section{Current conservation from the equation of motion}
\label{app:current_conservation}

In the main text, we obtained the U(1)-charge Noether-current conservation equation from our lattice Noether's theorem. Here, we show that this conservation equation can be derived from the EOMs (as in the continuum), either the one-step or the two-step one.
As above, we write only the indices that are shifted from $(j,p)$.

\subsection{From the one-step equation of motion}
From the (generic) one-step EOM, Eq.\ \eqref{eq:one-step scheme} with $\mathcal{W}$ as in Eq.\ \eqref{eq:walk_operator}, we immediately obtain the following four equalities
\begin{subequations}
\begin{align}
\psi^{\dag}\psi_{j+1} &= \psi^{\dag} \left( W_{-1} \psi_{p+1} + W_1 \psi_{p-1} + W_0 \psi \right) \label{eq:eqtwo} \\
\psi_{j+1}^{\dag} \psi &= \left( \psi_{p+1}^{\dag} W_{-1}^{\dag} +\psi_{p-1}^{\dag} W_{1}^{\dag}  + \psi^{\dag} W_{0}^{\dag}    \right) \psi \label{eq:eqone}\\
\psi_{j-1}^{\dag} \psi &= \left( \psi_{p-1}^{\dag} W_{-1} + \psi_{p+1}^{\dag} W_1 + \psi^{\dag} W_0 \right) \psi \label{eq:eqfour} \\
\psi^{\dag} \psi_{j-1} &= \psi^{\dag} \left( W_{-1}^{\dag} \psi_{p-1} + W_1^{\dag} \psi_{p+1} + W_0^{\dag} \psi \right) \label{eq:eqthree} \, .
\end{align}
\end{subequations}
Summing the first two  equations, Eqs.\ \eqref{eq:eqone} and \eqref{eq:eqtwo}, and substracting the last two, Eqs.\ \eqref{eq:eqthree} and \eqref{eq:eqfour}, we obtain after some simplifications and rearrangements the following current-conservation equation,
\begin{equation}
\label{eq:current_general}
d^L_{\mu} Q^{\mu} = 0 \, ,
\end{equation}
where the ``temporal'' and the ``spatial'' components of the current $Q$ are
\begin{subequations}
\begin{align}
    Q^0 &\defeq \frac{\epsilon}{2} \left(  \psi^{\dag} \psi_{j+1} + \psi^{\dag}_{j+1} \psi  \right) \\
    Q^1 &\defeq \frac{\epsilon}{2}\left(  \psi^{\dag} ( W_1^{\dag} - W_{-1}) \psi_{p+1} + \psi^{\dag}_{p+1} (W_1 - W_{-1}^{\dag} )\psi_{p} \right) \, . \label{eq:Q1current}
\end{align}
\end{subequations}

The temporal component of the current, $Q^0$, is actually equal to $J^0_{\text{U(1)}}$ defined in Eq.\ \eqref{eq:Jmu}.
Moreover, with the choices of Eqs.\ \eqref{eq:B} and \eqref{eq:choiceV}, we have that
\begin{subequations}
\begin{align}
    \mu_{\epsilon} \alpha^1 &= B_- \defeq W_1 - W_{-1} \\
     \mu_{\epsilon} &=B_+ \defeq W_1 + W_{-1} \, .
\end{align}
\end{subequations}
This implies
\begin{subequations}
\begin{align}
    W_1 &= \frac{ \mu_{\epsilon}}{2} (1 +\alpha^1) = W_1^{\dag} \\
    W_{-1} &= \frac{ \mu_{\epsilon}}{2} (1 -\alpha^1) = W_{-1}^{\dag} \, ,
\end{align}
\end{subequations}
so that
\begin{subequations}
\begin{align}
    W_1 - W_{-1}^{\dag} &=  \mu_{\epsilon} \alpha^1 \\
    W_1^{\dag} -W_{-1} &=  \mu_{\epsilon} \alpha^1 \, ,
\end{align}
\end{subequations}
which, inserted into Eq.\ \eqref{eq:Q1current}, yields $Q^1 = J^1_{\text{U(1)}}$, which ends our proof.

\subsection{Directly from the two-step equation of motion}

Multiplying the two-step EOM, Eq.\ \eqref{eq:DQW_fermions}, from the left by $2\mathrm{i}\psi^{\dag}$ we obtain
\begin{equation}
\label{eq:first_eq}
\psi^{\dag} \psi_{j+1} = \psi^{\dag} \psi_{j-1} - \psi^{\dag} \tilde{\alpha}^1 [\psi_{p+1} - \psi_{p-1}] - 2 \mathrm{i} m \psi^{\dag} \tilde{\alpha}^0 \psi \, .
\end{equation}
Similarly, taking the adjoint of Eq.\ \eqref{eq:DQW_fermions} and multiplying it from the right by $2\mathrm{i} \psi$ yields
\begin{equation}
\label{eq:second_eq}
\psi_{j+1}^{\dag} \psi = \psi^{\dag}_{j-1} \psi -   [\psi_{p+1}^{\dag} - \psi_{p-1}^{\dag}]\tilde{\alpha}^1 \psi + 2 \mathrm{i} m \psi^{\dag} \tilde{\alpha}^0 \psi \, .
\end{equation}
The sum of these Eqs.\ \eqref{eq:first_eq} and \eqref{eq:second_eq} gives the same current-conservation equation as that obtained from Noether's theorem, namely, Eq.\ \eqref{eq:one-step_current-conservation}.
Note that the above derivation has a direct parallel with the case of naive fermions.

\section{Continuum limit of the gauge-field action}
\label{app:gauge-field_action}

It is easy to show that
\begin{equation}
\label{eq:THE_eq}
    F_{\mu\nu} F^{\mu\nu} = \sum_l (-F_{0l}^2 - F_{l0}^2) + \sum_{k,l} F_{kl}^2 \, .
\end{equation}
Moreover, we have after a few lines of computation that, at lowest order in $\epsilon$,
\begin{equation}
\label{eq:SGtime}
    S_G^{\text{time}} = -\frac{1}{2} \sum_n \sum_l \epsilon^4 (-(F_{0l})_n^2) \, ,
\end{equation}
and, since $(F_{0l})^2 = (F_{l0})^2$, Eq.\ \eqref{eq:SGtime} can be rewritten as
\begin{equation}
    S_G^{\text{time}} = - \frac{1}{4} \sum_n \sum_l \epsilon^4 (- (F_{0l})_n^2 - (F_{l0})^2_n ) \, . \label{eq:SGtimefinal}
\end{equation}
We also have after a few lines of computation that, at lowest order $\epsilon$,
\begin{subequations}
\begin{align}
\label{eq:SGspace}
    S_G^{\text{space}} &= -\frac{1}{2} \sum_n \sum_{\substack{k,l \\ k<l}} \epsilon^4 (F_{kl})_n^2 \\
    &= -\frac{1}{4} \sum_n \sum_{\substack{k,l \\ k<l}} \epsilon^4 [(F_{kl})_n^2 + (F_{kl})_n^2] \\
     &= -\frac{1}{4} \sum_n  \epsilon^4 \Bigg( \sum_{\substack{k,l \\ k<l}} (F_{kl})_n^2 + \sum_{\substack{k,l \\ k<l}} (F_{lk})_n^2 \Bigg) \\
     &= -\frac{1}{4} \sum_n \epsilon^4 \Bigg( \sum_{\substack{k,l \\ k<l}} (F_{kl})_n^2 + \sum_{\substack{k,l \\ l<k}} (F_{kl})_n^2 \Bigg)  \\
     &= -\frac{1}{4} \sum_n \sum_{k,l} \epsilon^4 (F_{kl})_n^2 \, . \label{eq:SGspacefinal}
\end{align}
\end{subequations}
Summing Eqs.\ \eqref{eq:SGtimefinal} and \eqref{eq:SGspacefinal} and taking into account Eq.\ \eqref{eq:THE_eq}, we obtain the desired continuum limit, Eq.\ \eqref{eq:cont_limit}.

\end{document}